\documentclass[12pt]{iopart}
\usepackage{graphicx}
\usepackage{color}

\begin{document}

\title{Circular motion in NUT space-time}

\author{Paul I. Jefremov and Volker Perlick}

\address{ZARM, University of Bremen, 28359 Bremen, Germany. 
\\
Email:
paul.jefremow@zarm.uni-bremen.de, volker.perlick@zarm.uni-bremen.de}
\vspace{10pt}
\begin{indented}
\item[]August 2016
\end{indented}

\begin{abstract}
We consider circular motion in the NUT (Newman-Unti-Tamburino) space-time. 
Among other things, we determine the location of circular time-like geodesic 
orbits, in particular of the innermost stable circular orbit (ISCO) and of 
the marginally bound circular orbit. Moreover, we discuss the von Zeipel 
cylinders with respect to the stationary observers and with respect 
to the Zero Angular Momentum Observers (ZAMOs). We also 
investigate the relation of von Zeipel cylinders to inertial forces,
in particular in the ultra-relativistic limit. Finally, we generalise the 
construction of thick accretion tori (``Polish doughnuts'') which
are well known on the Schwarzschild or Kerr background to the case
of the NUT metric. We argue that, in principle, a NUT source could be 
distinguished from a Schwarzschild or Kerr source by observing 
the features of circular matter flows in its neighbourhood.  
\end{abstract}

%
%
%
%
%

\section{Introduction}
The Newman-Unti-Tamburino (NUT) metric is a solution to the
vacuum Einstein equation that generalises the Schwarzschild 
metric. Whereas the Schwarzschild metric depends only on one
parameter, $M$, which is to be interpreted as a mass, the NUT
metric involves a second parameter, $n$, which is called the NUT 
charge or the gravitomagnetic charge. The solution was found 
by Newman, Tamburino and Unti~\cite{NewmanTamburinoUnti1963} 
in 1963. Just as the the Schwarzschild solution, it may either be
joined to an interior perfect fluid solution (see  Bradley et al.
\cite{BradleyFodorGergelyMarklundPerjes1999}) or considered
as a black-hole solution. In the latter case, the region beyond the
horizon is isometric to Taub's cosmological vacuum 
solution~\cite{Taub1951}; therefore, the space-time that
results from extending the NUT space-time over the horizon
is known as the Taub-NUT solution.

Since the discovery of the NUT metric its physical relevance is 
a matter of debate. Here we have to face the problem
that solving Einstein's (vacuum) field equation only gives us a
local expression for the metric; there are different ways of how
to construct a global space-time from this local form of the metric 
and these different ways may lead to quite different physical interpretations.
Misner~\cite{Misner1963} has demonstrated how to construct 
a global NUT space-time that is regular everywhere outside 
the horizon and spherically symmetric. The price he had to pay
was that the time coordinate had to be made periodic, so in
Misner's version of the NUT space-time through each event
outside the horizon there is a closed time-like curve. If one 
is not willing to accept such a violation of causality, one may 
resort to a global version of the NUT metric suggested
by Bonnor~\cite{Bonnor1969}. In his interpretation there
is a singularity on the half-axis $\theta = \pi$ which may be 
interpereted as a rotating massless rod. Closed time-like curves 
exist only in a neighbourhood of this singularity which may 
be arbitrarily small, so the situation is not worse than in the 
space-time of a rapidly rotating straight string which is widely 
accepted as physically relevant.  As a generalisation of
Bonnor's construction, Manko and Ruiz~\cite{MankoRuiz2005} 
have demonstrated that one may introduce an additional 
parameter $C$ into the NUT metric that allows to distribute 
the singularity over both half-axes $\theta =0$ and $\theta = \pi$ 
in a symmetric or asymmetric way. Also, we mention the possibility
of superimposing two NUT charges of opposite sign to get a 
metric that is regular everywhere except between the two 
sources on the axis, see McGuire and Ruffini~\cite{McGuireRuffini1975}.

It is certainly fair to say that the properties of the NUT metric, in
any interpretation, are somewhat unusual if not pathological. 
In an often quoted article, Misner~\cite{Misner1967} called
it a ``counter-example to almost anything''. On the other
hand, in the last hundred years we have learned that exact
solutions to Einstein's (vacuum) field equation should be taken
seriously. The Schwarzschild ``singularity'' at 
$r=2M$ was considered as unphysical for many years; now
we have strong evidence that black holes really exist
and that the Schwarzschild metric (and more generally the Kerr 
metric) describes Nature correctly. Therefore, we believe that
also the NUT metric should be considered as a hypothetical
candidate for objects that exist in Nature. If one is willing
to accept this idea as a working hypothesis, one has to 
investigate the question of how a NUT source could be 
detected, i.e., by what observational
means we would know a NUT source if we see one. As for
all astrophysical objects, in principle there are two such
means: One could observe the influence on light (i.e., the
lensing features) or the influence on massive particles. The
lensing features of NUT objects have been comprehensively 
discussed by Nouri-Zonoz and Lynden-Bell~\cite{NouriLynden1997};
an analytical formula for the shadow of a NUT black hole, which 
is a particular lensing feature, is contained as a special case in the 
work of  Grenzebach et al.~\cite{Grenzebach14}. The motion of
massive particles in the NUT space-time was studied in a 
mathematically very elaborate paper by Kagramanova 
et al.~\cite{Kagramanova10}; moreover, 
Hackmann and L{\"a}mmerzahl~\cite{HackmannLaemmerzahl2012} 
calculated an upper bound
for the NUT charge of the Sun by analysing the motion of Mercury. 
However, as to particle motion around compact NUT sources, in
particular around NUT black holes, we believe that the question
of how to relate the mathematical results to observational
features has not yet been completely answered. It is the
purpose of this paper to fill this gap. For the sake of simplicity,
and because we believe that this case is particularly relevant, we 
restrict to circular particle motion.

The paper is organised as follows. In Section~\ref{sec:NUT}
we summarise some basic features of the NUT metric. In
particular, we outline the differences between Misner's and
Bonnor's interpretation and the relevance of the Manko-Ruiz 
parameter.   In Section~\ref{sec:geo} we summarise some 
results on circular geodesics in the NUT space-time that will be
used later. In Section~\ref{sec:Zeipel} we discuss the so-called
von Zeipel cylinders in the NUT space-time which, in our view,
are very useful for illustrating circular motion.  
Section~\ref{sec:inertial} gives a short summary of how 
the von Zeipel cylinders are related to inertial forces. Finally,
in Section~\ref{sec:doughnut} we discuss  a certain class
of thick accretion tori, known as Polish doughnuts, around a 
NUT source. These exact analytical solutions of the Euler
equation for a perfect fluid motion have been studied before
on Schwarzschild and Kerr space-times, but not, to the
best of our knowledge, on NUT space-time. We will see that
the Polish doughnuts around a NUT source are qualitatively
different from Polish doughnuts around a Schwarzschild or
Kerr source, so the observation of such accretion tori
allows, in principle, to discriminate between a NUT source
and a Schwarzschild or Kerr source.   
       
Our conventions are as follows. Throughout the paper we use
units making $c$ and $G$ equal to unity. Greek indices are
taking values $0,1,2,3$ and Einstein's summation convention is 
used. Our choice of signature is $(+---)$.


\section{NUT space-time}\label{sec:NUT}

The Newman-Unti-Tamburino (NUT) metric is a solution to the
vacuum Einstein equation that depends on two parameters,
a mass $M$ and a gravitomagnetic charge (NUT parameter) $n$.
It was discovered  by Newman, Tamburino and 
Unti~\cite{NewmanTamburinoUnti1963} in 1963,
a detailed discussion can be found, e.g., in the book by 
Griffiths and Podolsk{\'y}~\cite{GriffithsPodolsky}. In spherical
coordinates $(t,r, \theta , \phi )$ the NUT metric reads
\begin{equation}
\fl \qquad \quad
g _{\mu \nu} dx^{\mu} dx^{\nu}= 
\frac{\Delta}{\Sigma} \Big( \mathrm{d}t -2n(\cos \theta +C)\mathrm{d}\phi \Big) ^2 -
\frac{\Sigma}{\Delta}\mathrm{d}r^2 -\Sigma (\mathrm{d}\theta^2 +
\sin^2\theta \mathrm{d}\phi^2)
\label{NUT-metric}
\end{equation}
where 
\begin{eqnarray}\label{eq:DeltaSigma}
\Delta= r^2 -2Mr -n^2 \, , \quad 
\Sigma= r^2 +n^2 \, .
\end{eqnarray}
We have written the metric in the general form, also allowing for an
additional parameter $C$ which was introduced by Manko and 
Ruiz~\cite{MankoRuiz2005}. This parameter may take any real value.
The original version of the NUT metric is the metric (\ref{NUT-metric}) with
$C=-1$.  We will discuss the relevance of the parameter $C$
below. 

Whereas the Schwarzschild space-time has a curvature singularity at $r=0$,
in the NUT space-time with $n \neq 0$ the radius coordinate may range from 
$- \infty$ to $\infty$. (As an alternative, one may assume that the metric
(\ref{NUT-metric}) is valid only for $r_*<r$, with some radius value $r_*$, 
and that at $r= r_*$ it is matched to an interior metric, e.g. to a perfect 
fluid solution of Einstein's field equation, 
see Bradley et al. \cite{BradleyFodorGergelyMarklundPerjes1999}.
However, we will not consider this case here.) 
The metric coefficient $g_{rr}$ becomes infinite at
two radius values where $\Delta =0$. These are coordinate singularities
which indicate horizons. We shall restrict our consideration to the 
\emph{domain of outer communication}, i.e., to the region outside of the
outer horizon, 
\begin{eqnarray}\label{eq:r+}
r_{\mathrm{hor}} < r < \infty \, , \quad 
 r_{\mathrm{hor}} = M + \sqrt{M^2 +n^2} \, .
\end{eqnarray}
The adjacent region  
$M-\sqrt{M^2+n^2} < r < M+ \sqrt{M^2+n^2}$, which will be of no interest 
to us, is known as the Taub region because there the space-time is isometric to 
Taub's solution~\cite{Taub1951} to Einstein's vacuum field equation. 

The NUT metric can be generalised to the Kerr-NUT metric which involves,
in addition to $M$ and $n$, a specific angular momentum (Kerr parameter)
$a$. The Kerr-NUT metric reads (cf. Griffiths and Podolsk{\'y}~\cite{GriffithsPodolsky})
\begin{equation}
\fl \quad \;
g _{\mu \nu} dx^{\mu} dx^{\nu}= 
\frac{\Delta'}{\Sigma'}(\mathrm{d}t -\chi \mathrm{d}\phi)^2 -
\frac{\Sigma'}{\Delta'}\mathrm{d}r^2 -\Sigma' \mathrm{d}\theta^2 -
\frac{\sin^2\theta}{\Sigma'} \Big( a\mathrm{d}t-(\Sigma '  +a\chi)\mathrm{d}\phi \Big) ^2
\label{KNUT-metric}
\end{equation}
where $\Delta'=r^2 -2Mr +a^2 -n^2$, $\Sigma'= r^2+(n+a\cos\theta)^2$ 
and $\chi= a \sin^2\theta -2n(\cos \theta +C)$.
This metric with $M>0$ describes a black hole provided that $M^2+n^2 \ge a^2$. 
The outer horizon is at $r_{\mathrm{hor}} = M + \sqrt{M^2 +n^2-a^2}$. In this paper 
we will restrict to the NUT metric but we will occasionally refer, for the sake of 
comparison, to the Kerr metric, i.e., to the metric (\ref{KNUT-metric}) with $n=0$. 
A common feature of the NUT and of the Kerr metric is that they are stationary 
but non-static, because of the non-vanishing $g_{t \phi}$ term. A major difference is in the fact 
that the Kerr metric is invariant under a transformation $\theta \mapsto \pi - \theta$ whereas the 
NUT metric is not. Thus, the NUT metric is not symmetrical with respect to an equatorial plane 
as will be clearly seen by several features to be discussed below. 

We will now discuss the physical relevance of the Manko-Ruiz parameter $C$. Whereas the 
parameters $M$ and $n$ characterise invariant properties of the NUT metric, $C$ is to a certain 
extent  subject to coordinate transformations. First we observe that a coordinate transformation
\begin{eqnarray}\label{eq:Cpm}
\overline{t} = t \, , \quad \overline{\phi} = - \phi \, , \quad
\overline{\theta} = \pi - \theta \, , \quad \overline{r} = r
\end{eqnarray}
transforms $C$ to $\overline{C} = - C$. This demonstrates that NUT metrics with 
$C$ and $-C$ are globally isometric, where the isometry is given by a simultaneous 
inversion of the $\phi$ coordinate  and reflection at the equatorial plane. Moreover, a coordinate 
transformation 
\begin{eqnarray}\label{eq:CCp}
t' =  t - 2 n ( C-C') \phi  \, , \quad \phi' = \phi \, , \quad
\theta ' =  \theta \, , \quad r' = r
\end{eqnarray}
transforms a NUT metric with $C$ to a NUT metric with $C'$, for any real numbers 
$C$ and $C'$. However,  (\ref{eq:CCp}) is not a globally well defined
transformation unless we assume that $t$ is periodic with period $4 \big|n(C-C' ) \big| \pi$.
The reason is that $\phi$ is assumed to be periodic with period $2 \pi$. 
We may summarise these observations in the following way:
If we are not willing to make the time coordinate periodic, NUT space-times with 
Manko-Ruiz parameters $C$ and $C'$ are non-isometric and, thus, physically distinct
if $|C| \neq |C'|$. However, they are \emph{locally} isometric, as 
demonstrated by (\ref{eq:CCp}), on any neighbourhood that does not contain 
a complete $\phi$ line. If the time coordinate has period $T$, NUT space-times with 
Manko-Ruiz parameters $C$ and $C'$ are globally isometric if $4|n(C-C')| \pi = T$. 

The Manko-Ruiz parameter is of particular relevance in view of the fact that the NUT metric features
a singularity on the axis which is different from the trivial coordinate singularity of spherical
polars that is familiar even from flat space. We can read this from the contravariant time-time component of the metric which can easily be calculated from (\ref{NUT-metric}),
\begin{eqnarray}\label{eq:gtt}
g^{tt} = \frac{4n^2(\mathrm{cos} \, \theta + C )^2}{\Sigma \, \mathrm{sin} ^2 \theta}
+ \frac{\Sigma}{\Delta} \, .
\end{eqnarray}
While the second term is finite everywhere on the domain of outer communication, the first
one diverges for $\theta \to 0$  unless $C=-1$ and for $\theta \to \pi$ unless $C=1$.
There are two different ways of how to interpret this singularity, depending
on whether or not one is willing to make the time coordinate periodic. 

The first interpretation was brought forward by Misner~\cite{Misner1963}. He considered
the metric (\ref{NUT-metric}) with $C=-1$ which is regular except on the negative half-axis, 
$\theta = \pi$. By a coordinate transformation (\ref{eq:CCp}) with $C'-C=2$ he got a 
NUT metric with $C'=1$ which is regular except on the positive half-axis, $\theta = 0$.  
This transformation requires 
the time coordinate to be periodic with period $8|n| \pi$. Cutting out the singular half-axis 
from  each of the two copies
gives two coordinate patches that can be glued together to give a NUT space-time that is 
regular on both half-axes. This space-time is globally spherically symmetric, i.e.,
there is no distinguished axis. Of course, assuming 
that the time coordinate is periodic means that there is a closed time-like curve
through any event of the domain of outer communication, i.e., that the space-time 
violates the causality condition in an extreme way. However, one could argue that 
this is of no practical relevance if one assumes the period $T=8|n| \pi$ to be 
very \emph{large}. Misner's construction  can be generalised to gluing together 
a NUT metric with $C$ and a NUT metric with $C'=-C$, for any $C \neq 0$, provided that
the time coordinate is periodic with period $8|nC|\pi$. However, in the case $|C| \neq 1$ 
this is not very interesting because the resulting space-time still has a singularity
on the axis.

The second interpretation, which goes back to Bonnor~\cite{Bonnor1969},
is based on the assumption that the $t$ coordinate is \emph{not} periodic. 
Then (\ref{eq:CCp}) is not a globally allowed transformation and, for any choice of 
$C$, we have a true singularity on at least one half-axis. Bonnor interpreted this 
singularity as a massless spinning rod. He considered only the case that $C=-1$
where the singularity is on the negative half-axis. For
$C=1$ it is on the positive half-axis, for
$C=0$ it is distributed symmetrically and for any other value of $C$ 
asymmetrically on both half-axes. In Bonnor's version  the NUT metric 
is locally spherically symmetric, near any point off the axis, but not
globally in contrast to Misner's version with $|C|=1$. As we read from 
(\ref{NUT-metric}) that the $\phi$ lines are spacelike near the singularity,
also in Bonnor's interpretation the NUT metric contains closed time-like 
curves in the domain of outer communication. However, by choosing $|n|$ 
and $|C|$ sufficiently \emph{small} one may restrict them to a narrow 
neighbourhood of the axis so that they may be considered as irrelevant 
for possible motions of observers.    

\section{Time-like circular geodesics}\label{sec:geo}

It is well known that in the NUT space-time the Hamilton-Jacobi equation 
for the geodesics is separable which results in the following 
first-order equations for the time-like geodesics, cf.~Kagramanova et 
al.~\cite{Kagramanova10}:
\begin{eqnarray}
\label{eq:dr}
\dot{r}^2= R:= 
\frac{1}{4M^2}\left(\Sigma^2 E^2 -\Delta (r^2 +L^2 +K)\right),
\\
\label{eq:dtheta}
\dot{\theta}^2=\Theta:= 
\frac{1}{4M^2}\left(K +L^2 -n^2 -
\frac{\Big( L-2nE ( \cos\theta + C )\Big) ^2}{\sin^2 \theta}\right),
\\
\label{eq:dphi}
\dot{\phi}=
\frac{1}{2M}\left(\frac{L-2nE( \cos \theta + C )}{\sin^2 \theta}\right),
\\
\label{eq:dt}
\dot{t}=\frac{1}{2M}
\left(E\frac{\Sigma^2}{\Delta} 
+2n \Big( \cos \theta +C \Big) \, \frac{\Big( L-2nE( \cos \theta + C ) \Big)}{\sin^2 \theta}\right)
\, .
\end{eqnarray}
Here $E$ is the energy, $L$ is the $z$ component of the angular momentum
and $K$ is the Carter constant. (Note that Kagramanova et  al.~\cite{Kagramanova10} 
use a Carter constant $k$ which is related to our $K$ by $4M^2k=K$.) 
A dot denotes derivative 
with respect to the Mino time which is \emph{not} an affine parameter. 
By a coordinate transformation (\ref{eq:CCp}) with $C'=0$, which in any case
is well-defined locally near any point off the axis, one can get rid of the Manko-Ruiz 
parameter $C$ in (\ref{NUT-metric}). This transforms the equations for time-like
geodesics to
\begin{eqnarray}
\dot{r}^2=  
\frac{1}{4M^2}\left(\Sigma^2 E'^2 -\Delta (r^2 +L'^2 +K')\right),\\
\dot{\theta}^2= 
\frac{1}{4M^2}
\left(K' +L'^2 -n^2  -\frac{\Big( L' -2nE'\cos\theta \Big) ^2}{\sin^2 \theta}\right),\\
\dot{\phi}=
\frac{1}{2M}\left(\frac{L'-2nE' \cos \theta}{\sin^2 \theta}\right),\\
\dot{t'}=
\frac{1}{2M}
\left(E'\frac{\Sigma^2}{\Delta} +
2n\cos \theta \frac{\Big( L'-2nE' \cos \theta \Big) }{\sin^2 \theta}\right)
\end{eqnarray}
where the new constants of motion are related to the old ones from
(\ref{eq:dr}), (\ref{eq:dtheta}), (\ref{eq:dphi}) and (\ref{eq:dt}) 
by
\begin{eqnarray}\label{eq:ELK}
E'=E \, , \quad
L'=L -2nCE \, , \quad
K'=K+4nCEL-4n^2C^2E^2 \, .
\end{eqnarray}
We see that the values of the constants of motion
$L$ and $K$ assigned to a particular geodesic depend on $C$.
However, as curves in the space-time the time-like geodesics
are locally  independent of $C$ in the sense that, near any point
off the axis, they can be diffeomorphically mapped from  a
space-time with arbitrary $C$ onto a space-time with $C'=0$. As
this diffeomorphism involves only a transformation of the 
$t$ coordinate, the spatial
projections $\big( \phi (s) , \theta (s) , r(s) \big)$ of time-like
geodesics are globally independent of $C$. 

It is now our goal to determine
all circular time-like geodesics in this space-time. In particular we will
determine the location of the innermost stable circular orbit and 
of the marginally bound orbit. These results will be of importance 
for us later and we believe that they have not been given correctly  
before. 

As we said, we restrict to the domain of outer communication. It is 
our goal to find all circular time-like geodesics about the $z$-axis.
All other circular time-like geodesics then
follow immediately by applying all possible rotations. In space-times
with a singularity on the axis (i.e., in all cases except in Misner's 
spherically symmetric NUT space-time with $|C|=1$),  we
have to omit all curves that pass through this singularity. Note
that the spherical symmetry is broken only by the singularity on the 
axis: If this axis is removed, the hypersurfaces $r =\mathrm{constant}$
are homogeneous under the flow of a three-dimensional Lie algebra of
Killing vector fields that generate the rotation group, see the discussion 
in Bonnor's paper\cite{Bonnor1969} for the case $C=-1$. This flow
maps circular time-like geodesics onto circular time-like geodesics.  

Circular geodesics about the $z$ axis have to satisfy the four equations
$\dot{r}=0$, $\ddot{r}=0$, $\dot{\theta}=0$ and $\ddot{\theta}=0$. 
With the notation introduced in (\ref{eq:dr}) and (\ref{eq:dtheta}) 
these can be rewritten as 
\begin{eqnarray}\label{eq:RTheta}
R =0 \, , \quad
\frac{dR}{dr} =0 \, , \quad
\Theta =0 \, , \quad
\frac{d \Theta}{d \theta} =0 \, .
\end{eqnarray}
These four equations allow us to express the constants of motion
in terms of $r$ and $\theta$, 
\begin{equation}
E = \frac{\sqrt{\Sigma \Delta}
}{
\sqrt{\Sigma ^2-4n^2 \Delta \mathrm{tan} \, \theta}} \, ,
\label{EnCircNUT}
\end{equation}
\begin{equation}
L= \frac{\sqrt{4\Sigma \Delta} n(1+C \cos \theta)}{\cos \theta 
\sqrt{\Sigma^2 - 4n^2 \Delta \tan \theta}} \, ,
\label{DrehimpulsKreisNUT}
\end{equation}
\begin{equation}
K = n^2 - \frac{4 \Sigma \Delta n \Big( (1+C \, \mathrm{cos} \, \theta )^2
- \mathrm{sin} ^2 \theta \Big)}{\mathrm{cos} ^2 \theta \Big( \Sigma ^2
- 4n^2 \Delta \, \mathrm{tan} \, \theta \Big)} \, .
\label{Carter}
\end{equation}
Here we restrict to the case that $E>0$, i.e., that the tangent vector
to the time-like geodesic points into the same half of the light-cone 
as $\partial _t$. (Note that,  in contrast to the Kerr space-time, in the NUT
space-time there is no ergoregion, i.e., $\partial _t$ is time-like on the 
entire domain of outer communication.)    With $E$, $L$ and $K$
determined this way, (\ref{eq:RTheta}) leaves us with one
equation that gives us a relation between $r$ and
$\theta$,
\begin{eqnarray}\label{CosNUT}
\cos^2\theta = \frac{4n^2}{Q(r)}
\, ,
\end{eqnarray}
where
\begin{eqnarray}\label{eq:Q}
\fl \quad
Q(r) := \,  
\frac{6 n^6 r + 16 M^2 n^2 r^3 - 4 n^4 r^3 + 6 n^2 r^5 + 
 M (r^6 + 15 n^4 r^2 - 15 n^2 r^4 - n^6)}{r \Delta ^2} \, ,
\end{eqnarray}
i.e., it gives us the location of the circular time-like geodesics. 
(\ref{CosNUT}) reflects the fact that the NUT parameter
breaks the symmetry with respect to the equatorial plane: The circular 
time-like orbits are \emph{not} in the plane $\theta = \pi /2$ if 
$n \neq 0$. (This can also be read from (\ref{EnCircNUT}), 
(\ref{DrehimpulsKreisNUT}) and (\ref{Carter}).) They are rather 
located on two curved surfaces,  given by the
solution of (\ref{CosNUT}) with $\mathrm{cos} \, \theta >0$ and  
$\mathrm{cos} \, \theta <0$, respectively. On one surface the motion is
in the positive $\phi$ direction, on the other in the negative
$\phi$ direction.  
\footnote{Our result on the surfaces of 
circular motion is in disagreement with that of 
Chakraborty \cite{Chakraborty14}, who started his 
consideration of circular motion from the false 
assumption that the circular orbits lie in the 
plane $\theta =\frac{\pi}{2}$.} 

We have already observed that the projections of time-like 
geodesics to three-dimensional space are globally independent 
of the Manko-Ruiz parameter $C$. This is reflected by the fact 
that $C$ does not enter into equation (\ref{CosNUT}).

Circular time-like geodesics exist for radii from infinity down to the
radius of the photon sphere where the circular geodesic orbits become
light-like. In the NUT space-time, the radius of the photon sphere is 
given by 
\begin{equation}
4 r \Delta = \Sigma \frac{d \Delta}{dr} \, ,
\label{Equ-r_ph}
\end{equation}
cf. Grenzebach et al.~\cite{Grenzebach14}. The unique solution
outside the horizon is 
\begin{equation}
r_{\mathrm{ph}}= M + 
2 (M^2+n^2) ^{1/3} \, \mathrm{Re} \Big( (M+ i \, n) ^{1/3} \Big) 
\label{r_ph}
\end{equation}
which reduces to the well known result $r_{\mathrm{ph}}=3M$ in the 
Schwarzschild limit $n \to 0$. Note that in the NUT case the circular photon orbits 
are \emph{not} great circles, i.e., that for $r \to r_{\mathrm{ph}}$ the
angle $\theta$ as given by  (\ref{CosNUT}) does not go to $\pi /2$. 

However, not all the circular time-like geodesics between infinity and
the photon sphere are stable. In order to find the radius of the innermost 
stable circular orbit (ISCO) we have to consider again the equations
(\ref{EnCircNUT}), (\ref{DrehimpulsKreisNUT}), (\ref{Carter}) and 
(\ref{CosNUT}) for circular orbits, and we have to insert these 
expressions into the condition 
\begin{equation}
\frac{d^2 R}{d r^2} =0\, .
\label{ISCO-eq}
\end{equation}
which expresses marginal stability with respect to radial perturbations.
This leads to the equation 
\begin{eqnarray}
\fl \:
M r^6 - 6 M^2 r^5 - 15M n^2 r^4 + (4 M^2 n^2 - 16 n^4)r^3 + 15 M n^4 r^2 -6 M^2 n^4 r -M n^6=0
\label{ISCO-eqLast}
\end{eqnarray}
for the radius of the ISCO. This equation cannot be solved analytically
but it can be verified that there is a unique real solution 
$r=r_{\mathrm{ms}}>r_{\mathrm{hor}}$.
In the limit $n \to 0$ it gives the well-known ISCO radius in 
the Schwarzschild space-time, $r_{\mathrm{ms}}=6M$.
\footnote{The equation 
(\ref{ISCO-eqLast}) was obtained by Chakraborty \cite{Chakraborty14} 
as a limiting case (with $a \to 0$) of his equation for marginally 
stable orbits in the Kerr-NUT space-time. However, he incorrectly 
assumed throughout his work that the circular orbits, and hence the
ISCO, are in the equatorial plane $\theta = \pi/2$. Actually, the
$\theta$ coordinate of the ISCO is to be determined by inserting
the ISCO radius into (\ref{CosNUT}).}

Another important radius value is 
the radius of the marginally bound orbit which gives the lower bound 
for those orbits which do not go to infinity under small radial 
perturbations. From (\ref{EnCircNUT}) we read that $E \to 1$
if $r \to \infty$. Therefore, the marginally bound orbit is found
by solving the system of equations
\begin{equation}
R \big|_{E=1} =0\, , \quad
\frac{d R}{d r}\bigg|_{E=1} =0
\end{equation}
and is given by the expression
\begin{eqnarray}\label{eq:mb}
\fl \quad
r_{\mathrm{mb}} &=
M + \sqrt{M^2 + n^2 + \left( \frac{n^2 (M^2 + n^2)}{2 M} \right) ^{2/3}}
\nonumber
\\
\fl &+\sqrt{2 M^2 + 2 n^2 - \left( \frac{n^2 (M^2 + n^2)}{2 M} \right) ^{2/3}
+ \frac{2 M^4 + 3 M^2 n^2 + n^4}{
M \sqrt{M^2 + n^2 + \left( \frac{n^2 (M^2 + n^2)}{2 M} \right) ^{2/3}}}}
\, .
\end{eqnarray}
For $n \to 0$ this reduces to the well-known result $r_{\mathrm{mb}} = 4M$ for the
marginally bound orbit in the Schwarzschild space-time.
 
\begin{figure}
\begin{flushright}
\includegraphics[width=.8\linewidth]{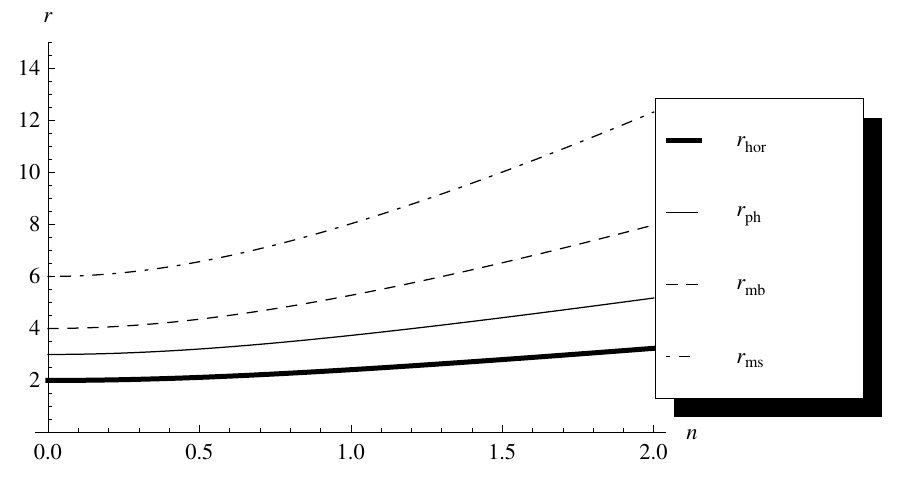}
\end{flushright}
\caption{Dependence on the NUT parameter $n$ of the 
radii of the horizon $r_{\mathrm{hor}}$, the photon 
sphere $r_{\mathrm{ph}}$, the marginally bound 
circular orbit $r_{\mathrm{mb}}$ and the 
marginally stable circular orbit (ISCO) $r_{\mathrm{ms}}$.
The values of $n$ and $r$ are given in units of $M$. }
\label{4Radii}
\end{figure}

\begin{figure}
\begin{flushright}
\includegraphics[width=.8\linewidth]{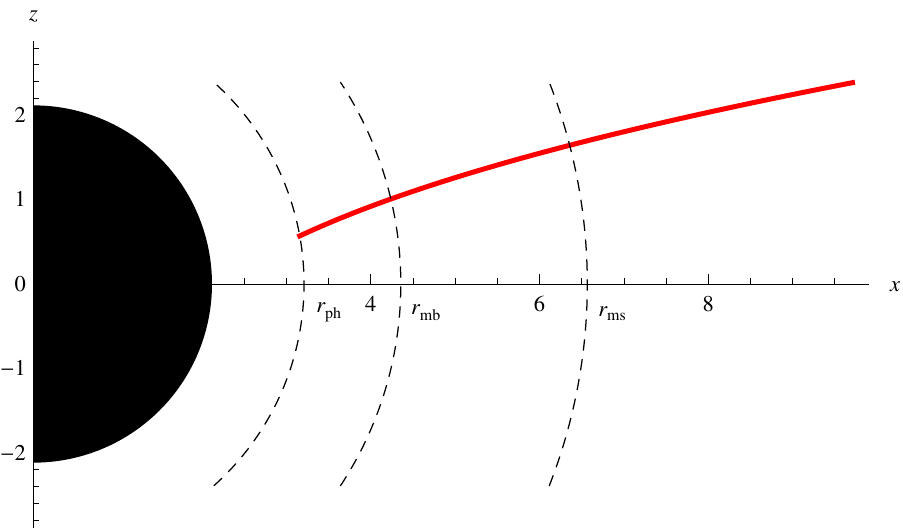}
\end{flushright}
\caption{Location of time-like circular geodesics in
the vicinity of a NUT black hole with $n=0.5 \, M$. On both axes $M$ 
is chosen as the unit.  One of the two
surfaces where geodesic circular motion about the $z$ axis is possible 
is marked by a thick (red) line. The other one is the mirror image with 
respect to the equatorial plane. Time-like circular geodesics exist only 
up to the photon sphere $r_{\mathrm{ph}}$; in the interval 
$r_{\mathrm{ph}}<r<r_{\mathrm{ms}}$ 
circular motion is unstable and for $r>r_{\mathrm{ms}}$ it is stable. The 
circular orbits with $r<r_{\mathrm{mb}}$ become unbound under radial
perturbations whereas for $r>r_{\mathrm{mb}}$ they remain bound, see the text
for further explanations.}
\label{CircOrb}
\end{figure}

The results on the photon sphere, the ISCO and the marginally bound
orbit in the NUT space-time are summarised in Fig.~\ref{4Radii}.
The location of the circular time-like 
geodesics is illustrated,
for the case $n = 0.5 \, M$, in  Fig.~\ref{CircOrb}.

\section{Von Zeipel cylinders}\label{sec:Zeipel}

In this section we want to study circular (in general non-geodesic)
particle motion in the NUT space-time and we want to use the so-called
von Zeipel cylinders for the sake of illustration. The relativistic notion
of von Zeipel cylinders was introduced by Abramowicz~\cite{Abramowicz71}
who showed that with the help of these cylinders a relativistic analogue
to the classical (i.e., non-relativistic) von Zeipel theorem can be
formulated. 

Before specifying to the NUT space-time, we re-establish some basic
facts about the von Zeipel cylinders that are valid in any 
axisymmetric and stationary space-time, so for the time being we
just assume that we have a metric of the form
\begin{equation}
g_{\mu \nu} dx^{\mu} dx^{\nu}=g_{tt}dt^2 +2g_{\phi  t}dt d{\phi} +g_{\phi \phi}d{\phi}^2 
+g_{rr}dr^2 +g_{\theta \theta} d{\theta}^2
\label{MetricStation}
\end{equation}
with two Killing vector fields $\partial_t$ and 
$\partial _{\phi}$. Note that there is a certain ambiguity in the
choice of $\partial _t$ and $\partial _{\phi}$ because any linear
combination with constant coefficients of two Killing vector fields
is again a Killing vector field. We restrict this ambiguity by requiring
that $\partial _t$ is time-like (on the entire domain of outer communication 
in the NUT space-time and outside of the ergoregion in the Kerr 
space-time) and that the $\phi$ coordinate is periodic with period
$2 \pi$. If the time coordinate is not periodic, this fixes $\partial _t$
up to an irrelevant constant factor and $\partial _{\phi}$ up to sign.
If the time coordinate is periodic with period $4 \big|n(C-C' ) \big| \pi$, 
we are still free to make coordinate transformations  (\ref{eq:CCp})
which leave the vector field $\partial _t$ invariant but not the 
hypersurfaces $t = \mathrm{constant}$.

In a space-time of the form (\ref{MetricStation}) where $\partial _t$ 
is time-like and $\phi$ is periodic with period $2 \pi$, circular motion 
about the $z$ axis is described by a four-velocity of the form
\begin{eqnarray}\label{eq:u}
u^{\mu} \partial _{\mu} =
\frac{\partial _t + \Omega \partial _{\phi}
}{\sqrt{g_{tt}+2 \Omega g_{t \phi}+ \Omega ^2 g_{\phi \phi}}}
\end{eqnarray}
where $\Omega$ is the angular velocity with respect to 
the stationary observers, i.e., observers moving on $t$ lines.
$\Omega$ may be chosen as an arbitrary function of $r$ and 
$\theta$, subject only to the condition that $\partial _t + \Omega
\partial _{\phi}$ be timelilke, 
\begin{eqnarray}\label{eq:time-like}
g_{tt}+2 \Omega g_{t \phi} + \Omega ^2 g_{\phi \phi} > 0 \, .
\end{eqnarray}
We may interpret $u^{\mu} \partial _{\mu}$ as the four-velocity 
of a fluid, but in this section we will restrict to purely kinematical 
considerations and not assume any equation of motion.

If we define the energy and the $z$ component of the angular
momentum in the usual way from the Lagrangian for free motion
as $E =  u_t$ and $L=-u_{\phi}$, the specific angular momentum
is given as
\begin{eqnarray}\label{eq:ell}
\ell = \frac{L}{E} =
\frac{- \Omega g_{\phi \phi}- g_{t \phi}}{g_{tt}+\Omega g_{t \phi}}
\, .
\end{eqnarray}
The motion can be characterised either by giving $\Omega$ as
a function of $r$ and $\theta$ or, equivalently, by giving
$\ell$ as a function of $r$ and $\theta$,
as (\ref{eq:ell}) can be solved for $\Omega$,
\begin{eqnarray}\label{eq:Omegaell}
\Omega =
\frac{\ell g_{tt} +  g_{t \phi}}{-g_{\phi \phi}- \ell g_{t \phi}}
\, .
\end{eqnarray}
If rewritten in terms of $\ell$, the condition (\ref{eq:time-like})
for time-like motion reads
\begin{eqnarray}\label{eq:elltime-like}
g_{\phi \phi} + 2g_{t\phi}\ell + g_{tt}\ell^2 < 0 \, .
\end{eqnarray}

The \emph{von Zeipel radius} (or \emph{radius of gyration}) 
$\mathcal{R}$ was defined by Abramowicz and collaborators
for the case of an axisymmetric and \emph{static}
metric (see in particular 
Abramowicz et al. \cite{AbramowiczMillerStuchlik1993})
by the definition
\begin{eqnarray}\label{Rgen}
\mathcal{R} ^2 = \frac{\ell}{\Omega} \, .
\end{eqnarray}
The surfaces $\mathcal{R} =\mathrm{constant}$ are known as
the \emph{von Zeipel cylinders}.  In the static case, $g_{t \phi} = 0$, 
inserting (\ref{eq:ell}) and (\ref{eq:Omegaell}) into (\ref{Rgen})
yields
\begin{eqnarray}\label{eq:Rstatic}
\mathcal{R} ^2 = \frac{-g_{\phi \phi}}{g_{tt}} \, .
\end{eqnarray}
On the domain where $\partial _t$ is time-like, the 
condition $g_{t \phi} =0$ implies that the right-hand 
side of (\ref{eq:Rstatic}) is non-negative
and $\mathcal{R}$ is well-defined as a non-negative universal 
function of $r$ and $\theta$, independent of $\Omega$. On the 
Schwarzschild space-time this is the case on the entire domain
of outer communication. 

Definition (\ref{Rgen}) makes sense also in the more general case of 
an axisymmetric and stationary metric with $g_{t \phi} \neq 0$, such as
the Kerr metric or the NUT metric. Then the von Zeipel radius 
$\mathcal{R}$ is given by the expression.
\begin{eqnarray}\label{R}
\mathcal{R} ^2 = \frac{\ell}{\Omega} =
\frac{-\Omega ^{-1} g_{t \phi} - g_{\phi \phi}}{g_{tt}+ \Omega g_{t \phi}} 
=
\frac{-\ell g_{t \phi} -g_{\phi \phi}}{\ell ^{-1} g_{t \phi}+g_{tt}}
\, .
\end{eqnarray}
The major difference in comparison to the case $g_{t \phi} = 0$ is in
the fact that now the function $\mathcal{R}$
depends on the motion. If the motion has been specified, by giving either 
$\Omega$ or $\ell$ as a function of $r$ and $\theta$, the von Zeipel radius 
is a well-defined function of $r$ and $\theta$ on the domain where the
right-hand side of (\ref{R}) is non-negative. This domain depends on the motion.
Note that in flat space-time
($g_{tt}=1, g_{t \phi} = 0 , g_{\phi \phi} =- r^2 \mathrm{sin} ^2 \theta$) 
we have $\mathcal{R} = r \, \mathrm{sin} \, \theta$, i.e., $\mathcal{R}$
is just the distance from the $z$ axis and the von Zeipel cylinders
are indeed Euclidean cylinders in the three-dimensional space.
As a consequence, if our axisymmetric and stationary space-time is 
asymptotically flat, the surfaces $\mathcal{R} = \mathrm{constant}$ have
the topology of cylinders if we are sufficiently far away from the axis.
In the inner region of the space-time, however, their topology
might be different.

For the NUT space-time, the relation (\ref{eq:ell}) between $\ell$ and $\Omega$
specifies to 
\begin{eqnarray}\label{ellNUT}
\ell= \frac{
\Omega \left( 4n^2(\mathrm{cos} \, \theta +C)-\frac{\Sigma ^2}{\Delta} \mathrm{sin} ^4 \theta \right)
}{
1- 2  \Omega  n (\mathrm{cos} \, \theta +C )
}
\end{eqnarray}
and the von Zeipel radius (\ref{R}) reads
\begin{eqnarray}\label{RNUT}
\mathcal{R} ^2 = \frac{
2 \ell n (\mathrm{cos} \, \theta +C) -4n^2 (\mathrm{cos} \, \theta +C)^2 
+ \frac{\Sigma ^2}{\Delta} \mathrm{sin} ^4 \theta 
}{
1 - 2 \ell^{-1} n (\mathrm{cos} \, \theta +C ) 
} \, .
\end{eqnarray}
With the help of these equations one easily verifies that
 the von Zeipel cylinders $\mathcal{R} = \mathrm{constant}$
are independent of the Manko-Ruiz parameter $C$ in the following sense: On a NUT 
space-time with $C \neq 0$, the von Zeipel cylinders for a motion with angular 
velocity $\Omega$ and specific angular momentum $\ell$ coincide with the 
von Zeipel cylinders on a NUT space-time with the same $M$ and $n$ and $C=0$
for a motion with
\begin{eqnarray}\label{eq:OmegaellC}
\Omega ' =  \frac{\Omega}{1-2nC \Omega} \, , \quad
\ell' = \ell -2nC \, .
\end{eqnarray}
We emphasise that the von Zeipel radius is a 
purely kinematic notion in the sense that it is well defined for any
circular motion without reference to an equation of motion. However,
it has particularly nice properties if we do assume a specific
equation of motion. For the case that the Euler equation of a perfect 
fluid with a barotropic equation of state is satisfied, 
Abramowicz \cite{Abramowicz71} has shown that the differential 
$d \Omega$ is a multiple of the differential $d \ell$ at all points where 
$d \ell \neq 0$. In other words, in this case $\Omega$ and thus $\mathcal{R}$ 
is constant on each non-degenerate surface $\ell =\mathrm{constant}$. 
This result is known as the \emph{relativistic von Zeipel
theorem}, cf. Rezzolla and Zanotti~\cite{Rezzolla13} for a detailed
discussion. 
 
Whereas (\ref{Rgen}) is the unanimously accepted definition of  the
von Zeipel radius in axisymmetric and static space-times, the generalisation
to the axisymmetric and stationary case is not unambiguous.  Abramowicz
et al.~\cite{Abramowicz95} argued that in the latter case
(with $g_{t \phi} \neq 0$) a modified von Zeipel
radius $\tilde{\mathcal{R}}$ should be introduced by the equation 
\begin{eqnarray}
\tilde{\mathcal{R}}{}^2= 
\frac{\tilde{\ell}}{\tilde{\Omega}}
= 
\frac{g_{\phi \phi}^2}{g_{t \phi}^2 -g_{tt}g_{\phi \phi}} \, ,
\label{R-tilde}
\end{eqnarray}
where $\tilde{\Omega}$ denotes the angular velocity with respect to the
zero angular momentum observers (ZAMOs), 
\begin{eqnarray}\label{eq:tOmega}
\tilde{\Omega}= \Omega - \omega \, , \quad
\omega = - \frac{g_{t \phi}}{g_{\phi \phi}} \, ,
\end{eqnarray}
and $\tilde{\ell}$ is the specific angular momentum with respect to the ZAMOs,
\begin{eqnarray}\label{eq:tell}
\tilde{\ell} = \frac{\tilde{L}}{\tilde{E}} = \frac{L}{E-\omega L} = \frac{\ell}{1-\omega \ell} \, .
\end{eqnarray}
The ZAMOs are observers 
whose worldlines are orthogonal to 
the hypersurfaces $t= \mathrm{constant}$. The name refers to the fact 
that these observers have $L=0$. In comparison to (\ref{R}) the
modified definition (\ref{R-tilde}) has the obvious advantage 
that $\tilde{\mathcal{R}}$ is a universal function of $r$ and
$\theta$, independent of the specific motion. Note that in the
Kerr-NUT space-time
$g_{t \phi} ^2 -g_{tt} g_{\phi \phi}$ is positive on the domain
of outer communication, so the right-hand side of (\ref{R-tilde}) is
non-negative there. However, when taking the square-root one has to
be careful with the sign of $g_{\phi \phi}$. In the Kerr space-time, 
$g_{\phi \phi}$ is negative on the entire domain of outer communication, 
but in the NUT space-time there is a region near the singularity on the 
axis, depending on $C$, where $\partial _{\phi}$ is time-like. As on this 
region the hypersurfaces $t = \mathrm{constant}$ are time-like, i.e.,
the ZAMOs would have to move at superluminal speed, it seems 
reasonable to omit this region. So we say that $\tilde{R}$ is defined
only on that part of the space-time where the hypersurfaces $t = \mathrm{constant}$
are spacelike and there we have
\begin{eqnarray}
\tilde{\mathcal{R}} 
= 
\frac{-g_{\phi \phi}}{\sqrt{g_{t \phi}^2 -g_{tt}g_{\phi \phi}}} \, .
\label{R-tilde2}
\end{eqnarray}

Abramowicz et al.~\cite{Abramowicz95} argue
that one should always work with $\tilde{\mathcal{R}}$
(which coincides, of course, with $\mathcal{R}$ in the case 
$g_{t \phi}=0$). Clearly, the fact that $\tilde{\mathcal{R}}$
is independent of the motion strongly supports this opinion.
On the other hand, $\mathcal{R}$ is of some relevance also
in the case $g_{t \phi} \neq 0$ because it is
associated with the above-mentioned relativistic von Zeipel 
theorem. Moreover, 
the following observation, which follows immediately from the
definitions, is of some importance: For a motion
with $\ell = \mathrm{constant}$, the differential $d \mathcal{R}$
is a multiple of $d \Omega$, so in
this case the surfaces $\mathcal{R} = \mathrm{constant}$
are surfaces of constant $\Omega$. An analogous 
statement is true if we replace $\ell$, $\Omega$ and $\mathcal{R}$
with $\tilde{\ell}$, $\tilde{\Omega}$ and $\tilde{\mathcal{R}}$, 
respectively. This indicates that both $\mathcal{R}$ and 
$\tilde{\mathcal{R}}$ are useful quantities.
\footnote{There are also other ways to define von Zeipel 
 cylinders, see e. \,g. \cite{Stuchlik13}.}
Henceforth we refer to the surfaces
$\mathcal{R} = \mathrm{constant}$ as to the ``von Zeipel cylinders
with respect to the stationary observers'' and to the surfaces
$\tilde{\mathcal{R}} = \mathrm{constant}$ as to the ``von Zeipel
cylinders with respect to the ZAMOs''. 

We now specify the function $\tilde{\mathcal{R}}$, which is 
independent of the specific motion, for the NUT metric. 
Inserting the metric coefficients into (\ref{R-tilde})
yields
\begin{equation}
\tilde{\mathcal{R}}{}^2= 
\frac{\Big( \Sigma^2 \sin^2 \theta -4 n^2 
\Delta (\cos \theta + C)^2 \Big) ^2}{\Delta \Sigma^2 \sin^2\theta}.
\label{eq:tRNUT}
\end{equation}
We see that $\tilde{\mathcal{R}}$ depends on the Manko-Ruiz parameter
$C$. The case $C=0$ plays a special role because in this case
 $\tilde{\mathcal{R}}{}^2$ is invariant under
a transformation $\theta \mapsto \pi - \theta$, i.e., the 
von Zeipel cylinders with respect to the ZAMOs are 
symmetric with respect to the equatorial plane.
For $n \to 0$ we get from (\ref{eq:tRNUT}) the well-known expression for the 
von Zeipel radius of the Schwarzschild metric,
\begin{eqnarray}\label{eq:RSchw}
\tilde{\mathcal{R}}{}^2 = \mathcal{R}{}^2 =
\frac{r^4 \mathrm{sin} ^2 \theta}{r^2-2Mr} \, .
\end{eqnarray}

\begin{figure}
\begin{flushright}
  \includegraphics[width=.4\linewidth]{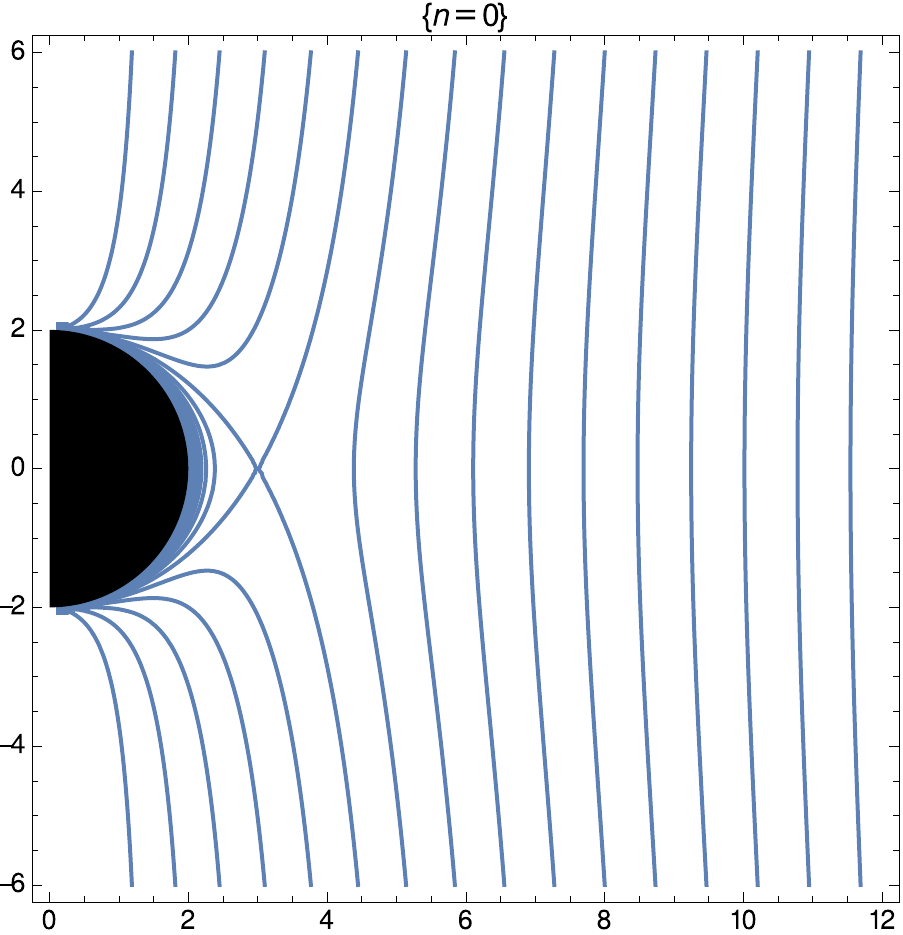}
\hspace{.01\linewidth}
  \includegraphics[width=.4\linewidth]{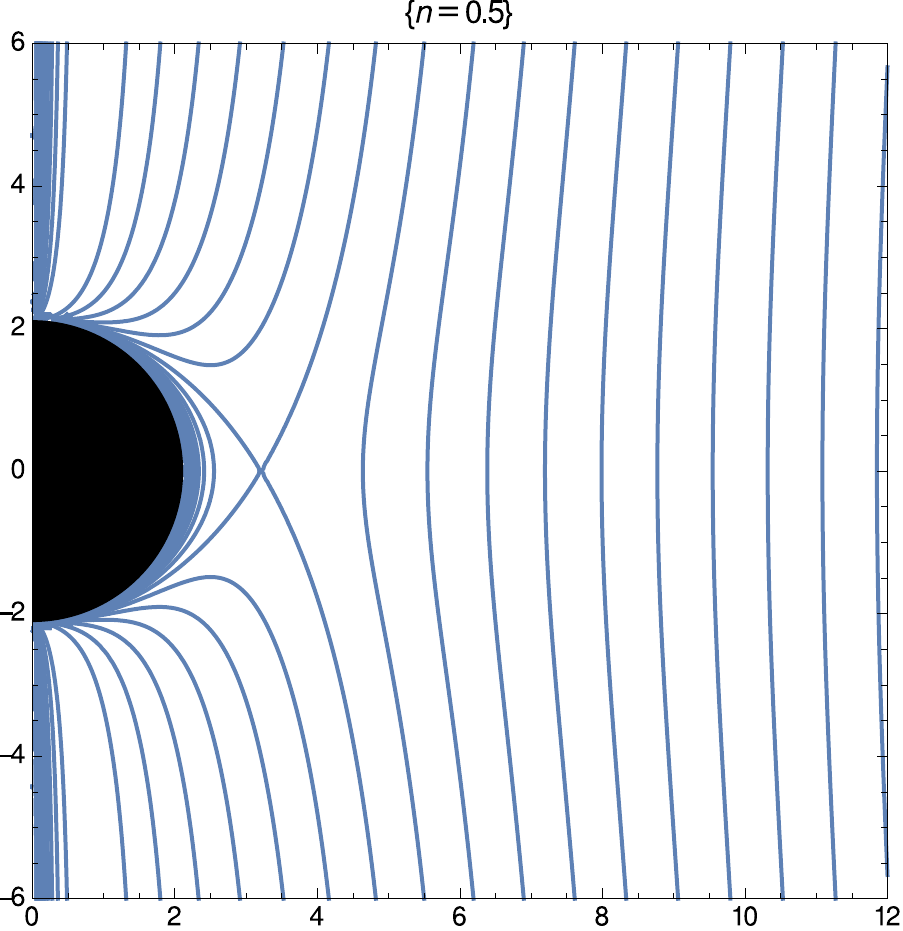}
\end{flushright}
\caption{Von Zeipel cylinders with respect to the ZAMOs, $\tilde{\mathcal{R}}
= \mathrm{constant}$,  in the NUT space-time with $n=0.5 \, M$ and $C=0$ (right)
and in the Schwarzschild space-time (left). 
In these pictures and in all the following ones, $M$ is used as the unit on the axes.}
\label{figR}
\end{figure}

\begin{figure}
\begin{flushright}
  \includegraphics[width=.4\linewidth]{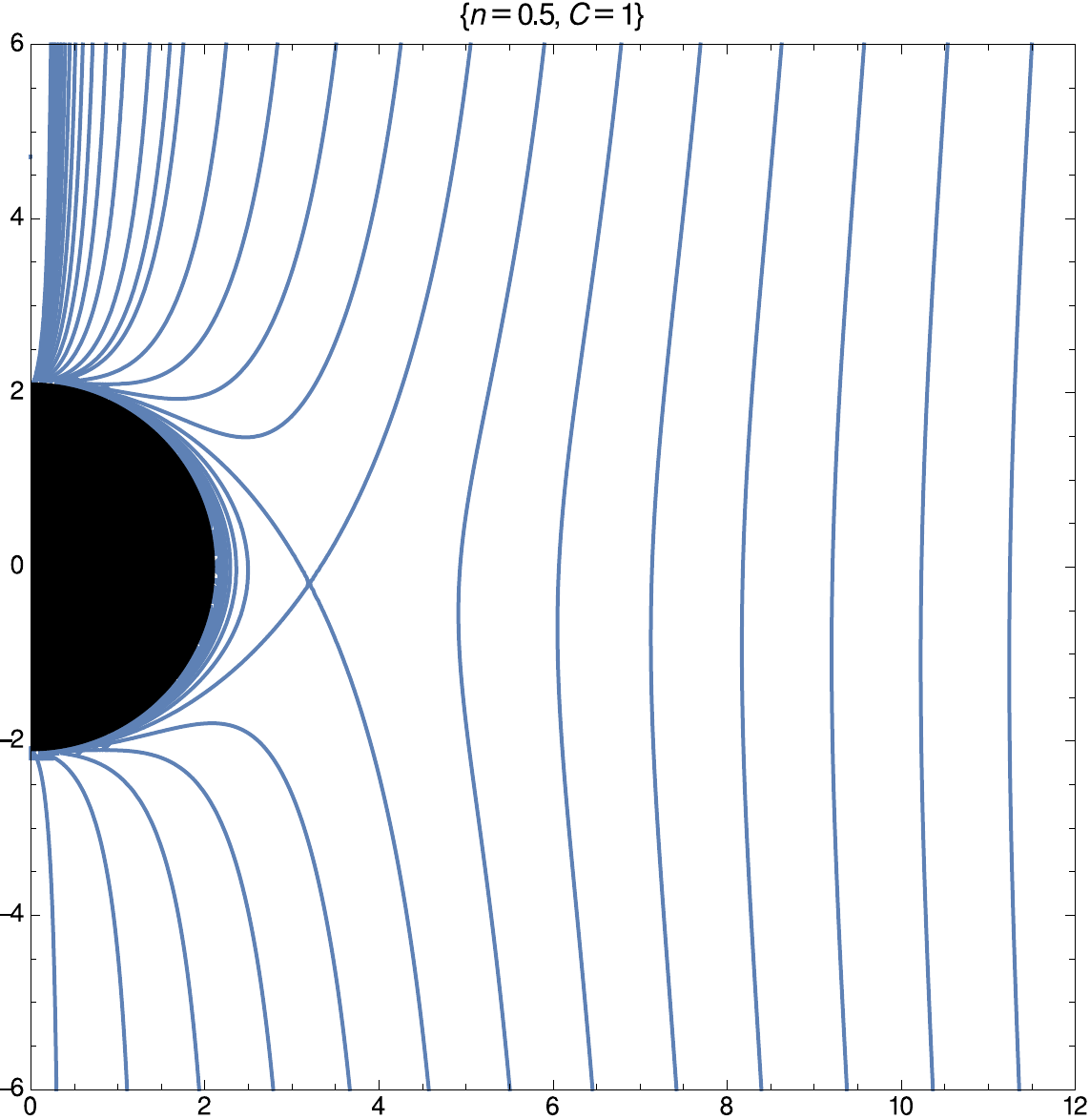}
\hspace{.01\linewidth}
  \includegraphics[width=.4\linewidth]{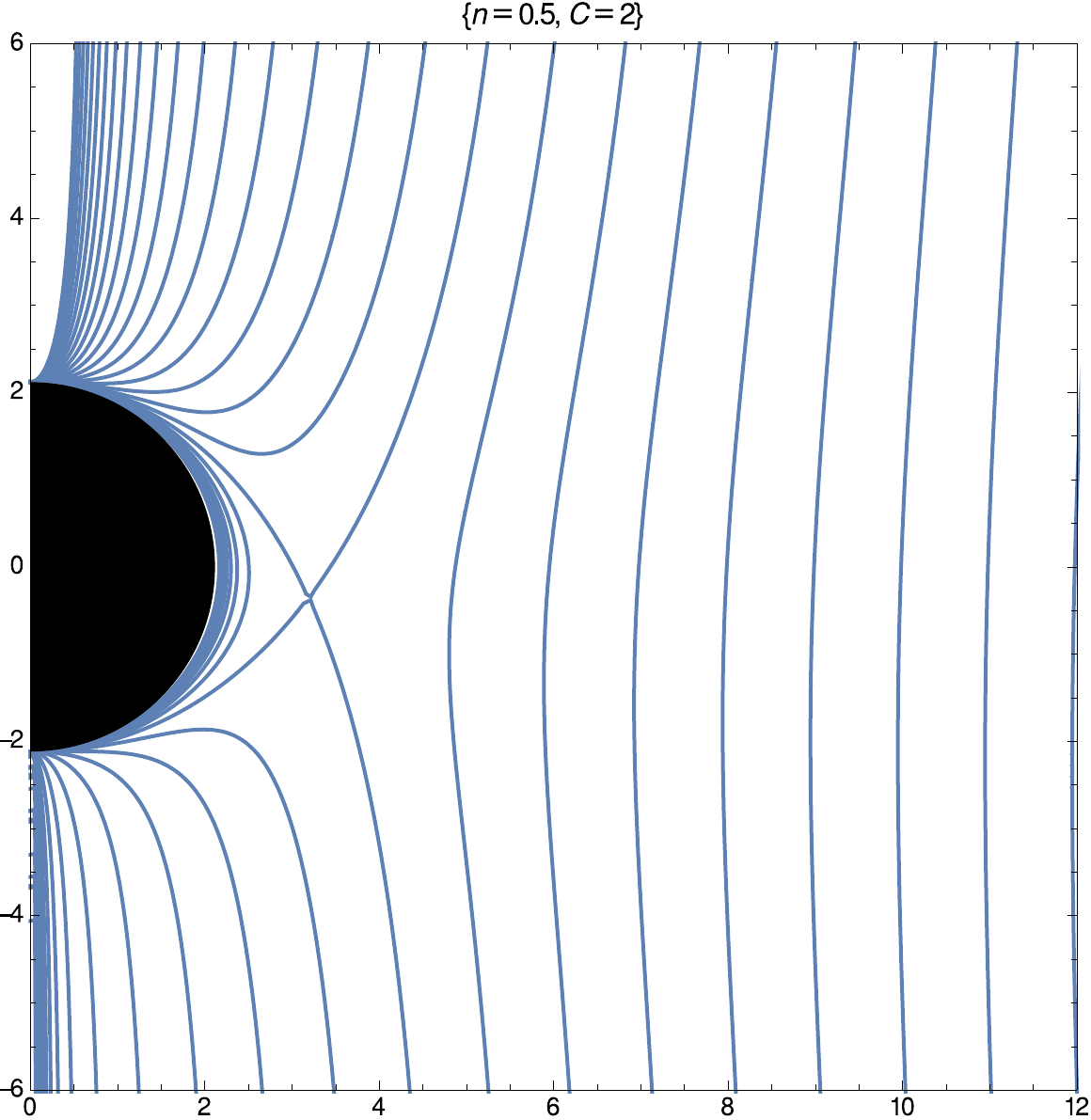}
\end{flushright}
\caption{Von Zeipel cylinders with respect to the ZAMOs, $\tilde{\mathcal{R}}
= \mathrm{constant}$,  in the NUT space-time with $n=0.5 \, M$ and $C=1$ (left)
and $C=2$ (right).}
\label{figR-2}
\end{figure}

Fig.~\ref{figR} illustrates the surfaces $\tilde{\mathcal{R}}
= \mathrm{constant}$ around a NUT black hole with $C=0$ and, for the 
sake of comparison, around a Schwarzschild black hole.
We see that they are qualitatively quite similar. In particular, in both
cases $\tilde{\mathcal{R}}$ is symmetric with respect to the 
equatorial plane. Note, however, that in the NUT case there
is a region near the singularity on the axis where $g_{\phi \phi} > 0$;
as already indicated above, we say that $\tilde{\mathcal{R}}$  is
not defined on this region. 
The situation is different for the NUT metric
with $C \neq 0$, see Fig.~\ref{figR-2}.   Then the symmetry
with respect to the equatorial plane is broken. The case $|C|=1$ 
is special because then only one half-axis is singular;
correspondingly, a region where $g_{\phi \phi} > 0$ 
exists only near one half-axis. Here we consider Bonnor's 
interpretation of the NUT metric. With Misner's interpretation, the
situation is more subtle. If the time coordinate is periodic, we may
transform the hypersurfaces $t= \mathrm{constant}$, and
thereby the ZAMOs,  according to (\ref{eq:CCp}). In Misner's
NUT space-time with $C=\pm1$, which is regular on both half-axes,
we have two families of ZAMOs and, correspondingly, two different
functions $\tilde{\mathcal{R}}$. For either family of ZAMOs the
domain of definition excludes a region near one of the two half-axes.
As in this case the space-time is spherically symmetric, we may choose
\emph{any} axis for this construction, so there are infinitely
many families of ZAMOs.

\begin{figure}
\begin{flushright}
  \includegraphics[width=.4\linewidth]{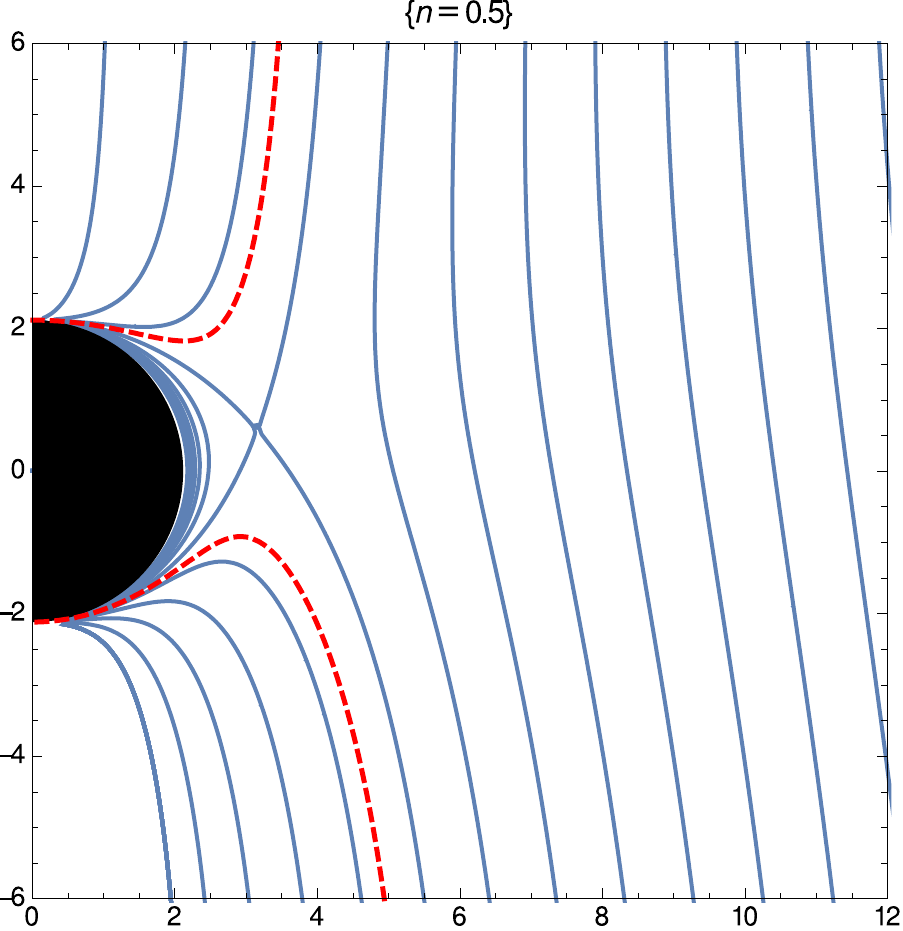}
\hspace{.01\linewidth}
  \includegraphics[width=.4\linewidth]{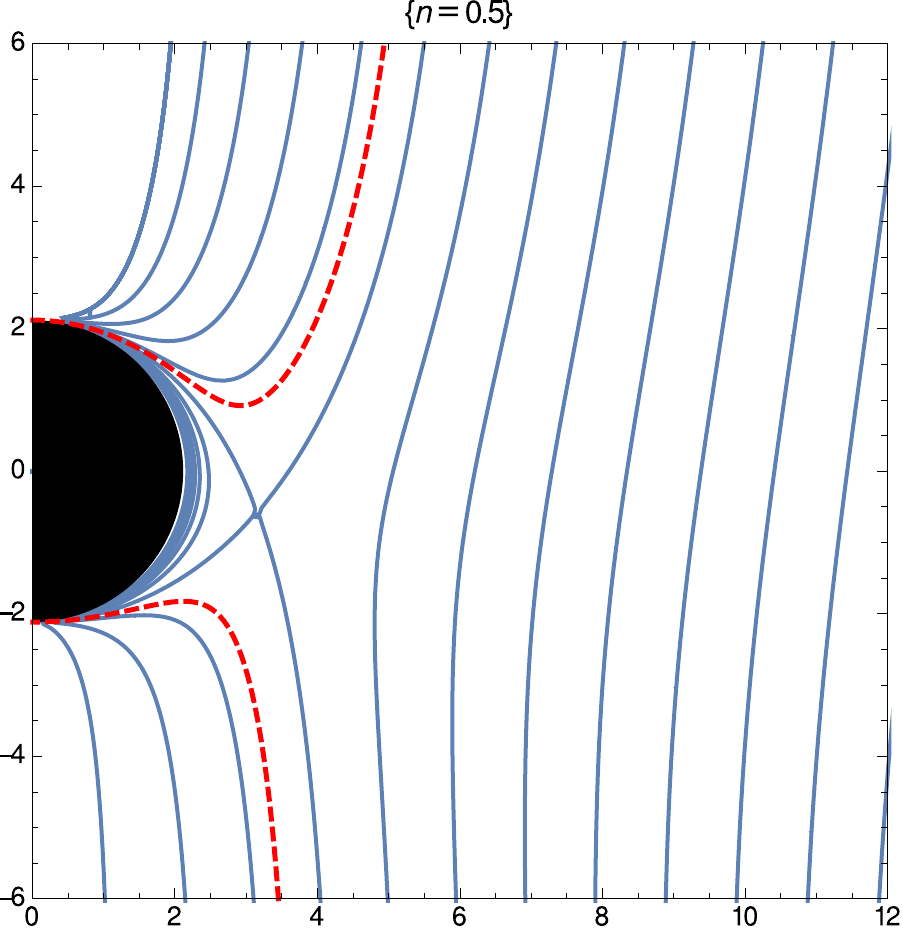}
\end{flushright}
\caption{Von Zeipel cylinders with respect to the stationary observers,
$\mathcal{R} = \mathrm{constant}$, 
in the NUT space-time with $n=0.5 \, M$ and $C=0$, for circular motion with 
$\ell = 5 \, M$ (left) and $\ell = - 5 \, M$ (right).
The (red) dashed lines denote the surfaces where the fluid velocity becomes
 light-like; this is the boundary of the region where circular time-like motion 
with the chosen $\ell$ is possible.
Reversing the sign of $\ell$ has the same effect as reversing the $z$-axis.}
\label{figR-Dyn2}
\end{figure}

\begin{figure}
\begin{flushright}
  \includegraphics[width=.4\linewidth]{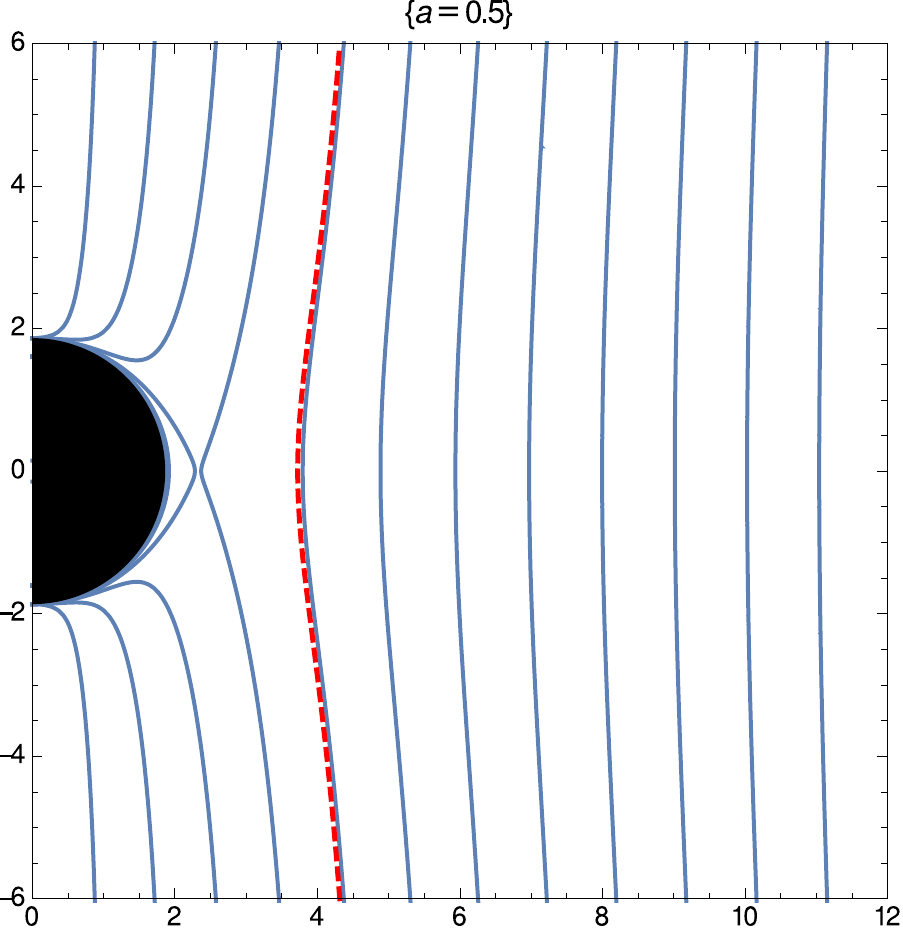}
\hspace{.01\linewidth}
  \includegraphics[width=.4\linewidth]{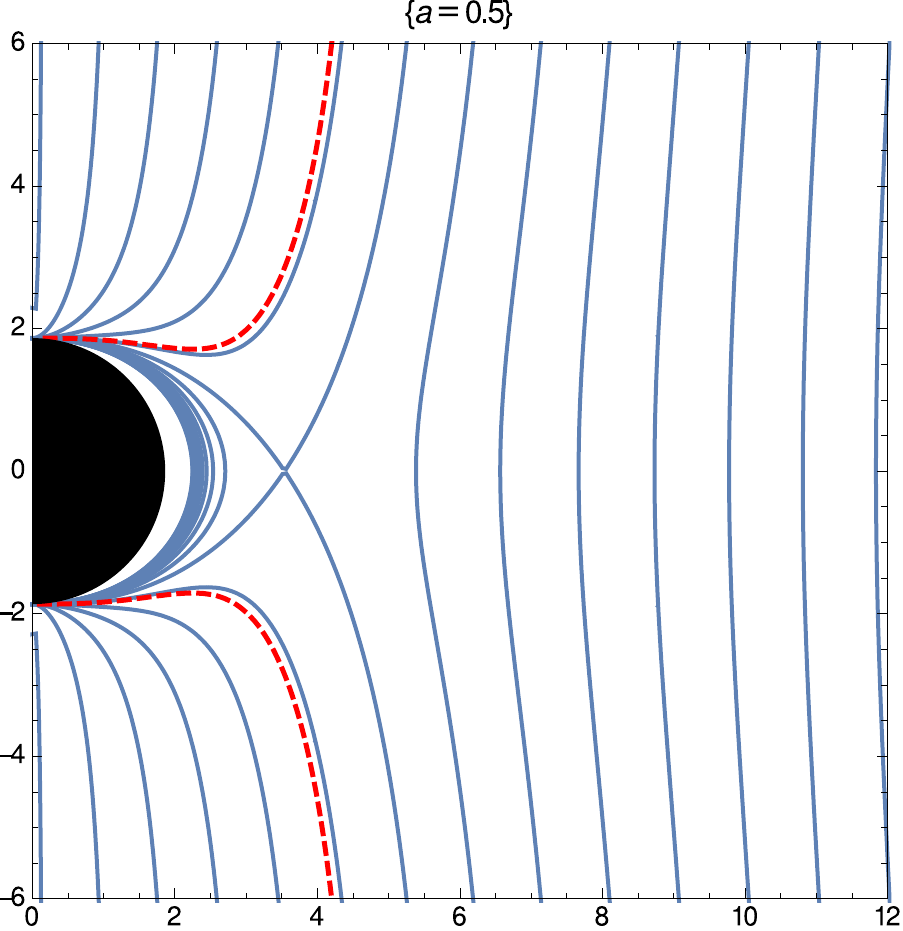}
\end{flushright}
\caption{Von Zeipel cylinders with respect to the stationary observers, $\mathcal{R}
= \mathrm{constant}$, in the Kerr space-time with 
$a=0.5 \, M$, for circular motion with $\ell = 5 \, M$ (left) and $\ell = - 5 \, M$ (right).
The (red) dashed lines have the same meaning as in Fig.~\ref{figR-Dyn1}.
As always, $M$ is used as the unit on the axes. }
\label{figR-Dyn1}
\end{figure}

We now turn to the surfaces $\mathcal{R} =
\mathrm{constant}$. In Figs.~\ref{figR-Dyn2} we plot these 
surfaces   in the NUT space-time and we compare
them with those in the Kerr space-time which are shown in
Fig.~\ref{figR-Dyn1}. As the surfaces $\mathcal{R} =
\mathrm{constant}$ are not universal, we have to specify the motion. This can be done 
by prescribing $\Omega$ or $\ell$ as a function of $r$ and $\theta$.
We have chosen a motion with $\ell = \mathrm{constant}$ which is the case
we will also consider in Section~\ref{sec:doughnut} below, so the surfaces
$\mathcal{R} = \mathrm{constant}$ coincide with the surfaces $\Omega
= \mathrm{constant}$. The
Manko-Ruiz parameter is set equal to zero which is no restriction
because we can reduce to this case by a transformation 
(\ref{eq:OmegaellC}). 
We see that in the NUT space-time the von Zeipel cylinders with respect to
the stationary observers are not symmetric with respect to the 
equatorial plane. Switching from $\ell$ to $-\ell$ has the same effect as
changing $\theta$ into $\pi - \theta$. We have indicated by a (red) dashed line the
boundary of the region where motion with the chosen $\ell$ is time-like, i.e., 
where the inequality (\ref{eq:time-like}) is satisfied. Formally, the function 
$\mathcal{R}$ is still well-defined outside of this region, because the 
right-hand side of (\ref{R-tilde})  is still positive, but it is not associated with a
physically reasonable motion; the right-hand side of (\ref{R-tilde})  becomes negative
only on a region near the half-axis $\theta = \pi$ for positive $\ell /n$ and
near the half-axis $\theta =0$ for negative $\ell /n$. The situation is completely 
different in the Kerr metric, see Fig.~\ref{figR-Dyn1}. Here the von Zeipel
cylinders with respect to the stationary observers are symmetric with
respect to the equatorial plane. Changing the sign of $\ell$ does \emph{not}
correspond to reversing the z axis; clearly, the reason is that in the Kerr case
the relative sign of $a$ and $\ell$ plays a role. Again, motion with the chosen
$\ell$ is time-like only on part of the domain of outer communication, which is
indicated by the (red) dashed line. In the Kerr case, the funtion $\mathcal{R}$ 
is well-defined, i.e., the right-hand side of (\ref{R-tilde})  is positive, except for the co-rotating 
case ($\ell \, a > 0$) on a certain region near the horizon.

For our purpose, the most important property of the von Zeipel cylinders is in
the fact that, for motions with $\ell = \mathrm{constant}$, the surfaces
$\mathcal{R} = \mathrm{constant}$ coincide with the surfaces $\Omega =
\mathrm{constant}$. In addition,
the von Zeipel cylinders are also related to inertial forces, as we will briefly
discuss in the following section.

\section{Inertial forces}\label{sec:inertial}
In this section we want to discuss, for the kind of circular motion
considered in the preceding section, the decomposition of the
four-acceleration in terms of inertial accelerations (or
inertial forces if we multiply with the mass of each particle).
In particular, we want to discuss the relation of the von Zeipel 
cylinders to the inertial forces. As in the 
preceding section, we begin with some general results which
hold for any axisymmetric and stationary metric before 
specifying the results for the NUT metric.   

There are various ways in which the decomposition of the 
four-acceleration can be introduced. In any case, this 
decomposition depends on the choice of a kind of reference 
frame. This can be done on an arbitrary space-time, see Foertsch et 
al.~\cite{Foertsch03}, but here we are interested only in 
axisymmetric and stationary space-times of the form 
(\ref{MetricStation}) where it is most convenient to
introduce the tetrad
\begin{eqnarray}
\fl \qquad
e_0=\frac{\partial_t +\omega \partial_{\phi}}{\sqrt{g_{tt}+\omega g_{t \phi}}}
\, , \quad
e_1= \frac{1}{\sqrt{-g_{\phi \phi}}}\partial_{\phi}
\, , \quad
e_2= \frac{1}{\sqrt{-g_{\theta \theta}}}\partial_{\theta}
\,, \quad
e_3= \frac{1}{\sqrt{-g_{rr}}}\partial_r \, ,
\label{ZAMO}
\end{eqnarray}
with $\omega$ being defined in (\ref{eq:tOmega}). The integral
curves of $e_0$ are the worldlines of the ZAMOs already considered
in the preceding section; the tetrad (\ref{ZAMO}) is also known as
the locally non-rotating frame (LNRF). We will decompose, following 
Abramowicz et al.~\cite{Abramowicz95}, the four-acceleration 
of particles in circular motion with respect to this tetrad. 

We begin by rewriting the four-velocity (\ref{eq:u}) of our 
particles as
\begin{equation}\label{eq:uv}
u= \frac{e_0 + v e_1}{\sqrt{1-v^2}},
\end{equation}
where $v$ is a function of $r$ and $\theta$ that gives, at each point,
the three-velocity as measured by a ZAMO. We distinguish between the 
four-acceleration as a vector, $a$, and the four-acceleration as a
covector, $A$, which are related by 
\begin{eqnarray}\label{eq:A}
A = g \big( a ,  \, \cdot \,  \big) \, , \quad
a = \nabla _u u \, .
\end{eqnarray}
The inertial acceleration is the acceleration of freely 
falling particles relative to our particles, i.e., it is given
by the covector $- A$. From (\ref{eq:uv}) and (\ref{eq:A}) we find that
the inertial acceleration can be decomposed in the form
\begin{eqnarray}\label{eq:Adec}
- A = A_{\mathrm{gr}} + A_{\mathrm{Cor}}+ A_{\mathrm{cen}}
\end{eqnarray}
where the first term is independent of $v$, the second term
is odd in $v$ and the third term is even in $v$,
\begin{eqnarray}
\label{eq:A1}
A_{\mathrm{gr}}=
-\frac{1}{2} \mathrm{d} 
\ln(-g_{tt}-\omega) \, ,
\\
\label{eq:A2}
A_{\mathrm{Cor}}= 
-\frac{v}{1-v^2} 
\tilde{\mathcal{R}}{} \mathrm{d}\omega \, ,
\\
\label{eq:A3}
A_{\mathrm{cen}}= 
\frac{v^2}{(1-v^2)} 
\tilde{\mathcal{R}}{}^{-1} \mathrm{d}\tilde{\mathcal{R}}{} \, 
\end{eqnarray}
cf. Abramowicz et al.~\cite{Abramowicz95}. $A_{\mathrm{gr}}$, 
$A_{\mathrm{Cor}}$ and $A_{\mathrm{cen}}$ are called the 
gravitational acceleration, the Coriolis acceleration and
the centrifugal acceleration, respectively. Note that each
of the three expressions (\ref{eq:A1}), (\ref{eq:A2}) and
(\ref{eq:A3}) is proportional to a differential, i.e., that
$a_{\mathrm{gr}}$,  $a_{\mathrm{Cor}}$
and $a_{\mathrm{cen}}$ are each perpendicular to a family 
of potential surfaces. 

Recall that in general $\mathcal{R}$ depends on $v$ whereas 
$\tilde{\mathcal{R}}$ does not. We will now investigate how 
$\mathcal{R}$ behaves in the ultrarelativistic limit, i. \,e. 
for $v \to \pm 1$. If we epxress $\Omega$ in terms of $v$, 
\begin{eqnarray}\label{eq:Omegav}
\Omega = \omega - 
\frac{\sqrt{g_{t \phi}^2-g_{tt}g_{\phi \phi} }}{g_{\phi \phi}}
 \, v = \omega + \frac{v}{\tilde{\mathcal{R}}}
\, ,
\end{eqnarray}
we find after a straight-forward calculation that the two definitions 
of von Zeipel radii (\ref{R}) and (\ref{R-tilde}) are related by
\begin{equation}
\mathcal{R}_{\pm}:=
\lim_{v \to \pm 1}\mathcal{R}
=\frac{\tilde{\mathcal{R}}{}}{1 \pm\omega \tilde{\mathcal{R}}{}} \, .
\label{R-lim}
\end{equation}
If we compare the expression
\begin{eqnarray}\label{eq:dtR}
d \mathcal{R}{}_{\pm} =
\frac{d \tilde{\mathcal{R}} \mp \tilde{\mathcal{R}}{}^2 d \omega
}{
\big( 1 \pm \omega \tilde{\mathcal{R}} \big) ^2}
\end{eqnarray}
with the fact that
\begin{eqnarray}\label{eq:ACc}
A_{\mathrm{Cor}}+ A_{\mathrm{cen}} =
\frac{v^2Z(v)}{(1-v^2) \tilde{\mathcal{R}}} \, , \quad
Z(v) = 
d \tilde{\mathcal{R}} - \frac{1}{v} \, \tilde{\mathcal{R}}{}^2 d \omega
\, , 
\end{eqnarray}
we see that in the ultrarelativistic limit the \emph{sum} of
Coriolis and centrifugal acceleration becomes perpendicular
to the potential surfaces $\mathcal{R}{}_{\pm} = \mathrm{constant}$,
cf. Hasse and Perlick~\cite{HassePerlick2006} where this result
was already found for the special case of the Kerr-Newman space-time.
If $g_{t \phi} = 0$ we have of course $\mathcal{R}{}_{\pm} = 
\tilde{\mathcal{R}} = \mathcal{R}$ and
$a_{\mathrm{Cor}}+a_{\mathrm{cen}}= a_{\mathrm{cen}}$ 
is perpendicular to the 
surfaces $\mathcal{R} = \mathrm{constant}$ for any $v$.

\begin{figure}
\begin{flushright}
  \includegraphics[width=.4\linewidth]{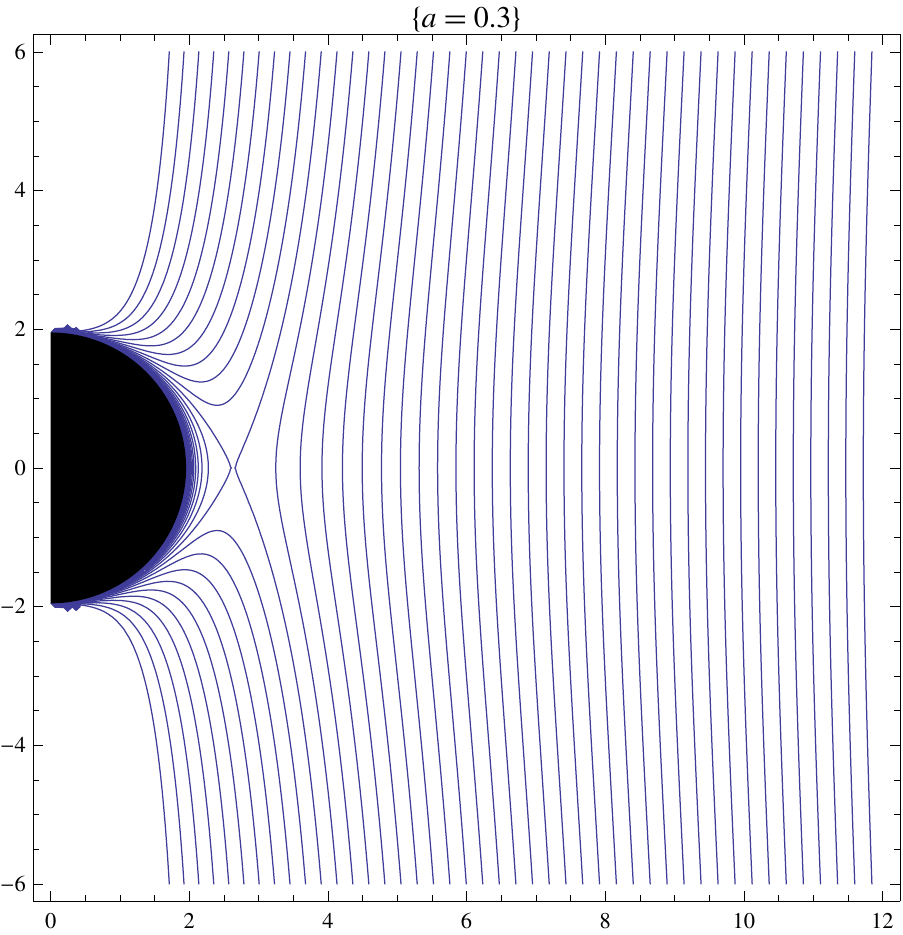}
\hspace{.01\linewidth}
  \includegraphics[width=.4\linewidth]{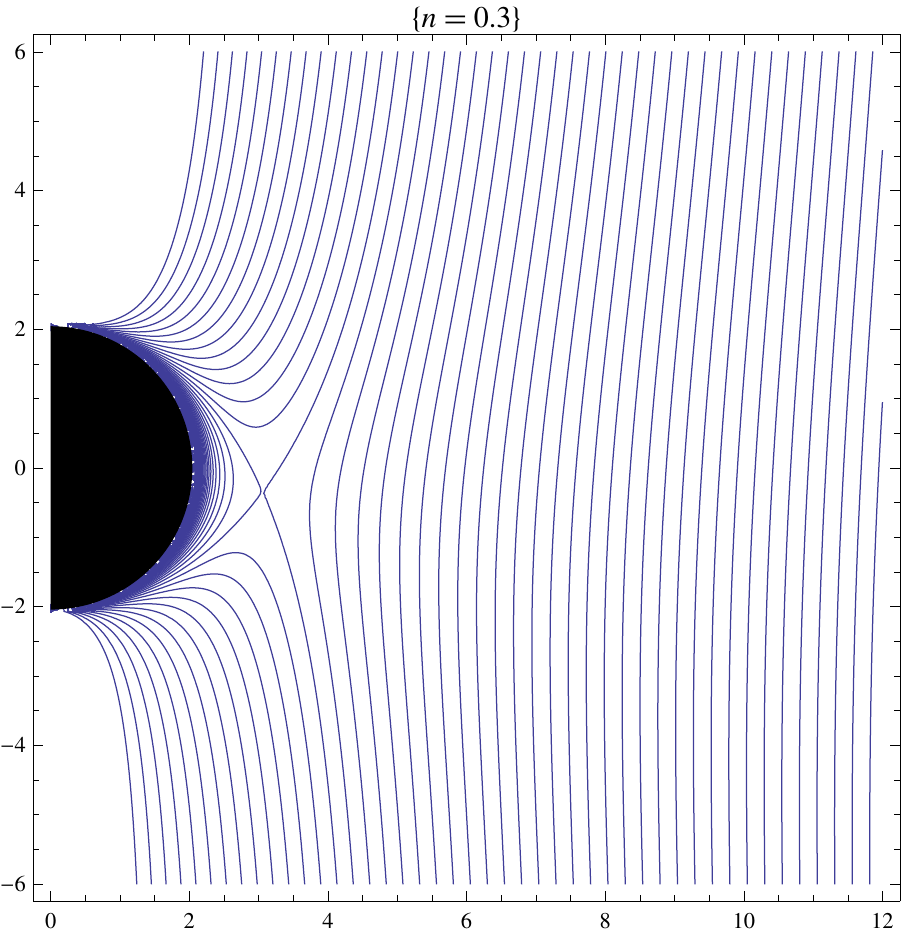}
\end{flushright}
\caption{The potential surfaces for the ultrarelativistic limit
of centrifugal-plus-Coriolis force $\mathcal{R}_{+}=\mathrm{constant}$ 
in Kerr space-time with $a=0.3 \, M$ (left) and in NUT space-time 
with $n=0.3 \, M$ (right).}
\label{figEP}
\end{figure}

In the NUT space-time the potential $\mathcal{R}{}_{\pm}$
is given by
\begin{eqnarray}
\mathcal{R}_{\pm} =  \frac{\big( r^2 + n^2 \big) \mathrm{sin} \, \theta}{\sqrt{r^2-2Mr-n^2}} \,
 \pm 2 n  \big( \mathrm{cos} \, \theta + C \big)  \, .
\label{R-limNUT}
\end{eqnarray}
As $C$ enters only in terms of an additive constant, the set of surfaces $\mathcal{R}{}_{\pm}
= \mathrm{constant}$ is independent of $C$. For the co-rotating case, i.e., for the 
plus sign, these  surfaces are plotted in Fig. \ref{figEP}, in comparison with the  Kerr metric.
In the NUT and in the Kerr case, there is exactly one value $(r, \theta)$ where 
$d \mathcal{R}{}_{\pm}=0$. This indicates a circular light-like geodesic, cf.  Fig.~\ref{CircOrb}.

\section{Polish doughnuts with constant specific angular momentum 
in NUT space-time}\label{sec:doughnut}
The so-called Polish doughnuts are analytical solutions 
for stationary fluid flows orbiting a black hole. They
have been introduced by Jaroszy{\'n}ski, Abramowicz and 
Paczy{\'n}ski~\cite{JaroszynskiAbramowiczPaczynski1980}
on Schwarzschild and on Kerr space-time; a detailed discussion
can be found, e.g., in the book by Rezzolla and 
Zanotti~\cite{Rezzolla13}. Here we want to discuss
Polish doughnuts on the NUT space-time to see 
how their shape is influenced by the NUT parameter.

We consider circular fluid flows, i.e., fluid flows with
a four-velocity of the form (\ref{eq:u}) with $\Omega$
being a function of $r$ and $\theta$. We asume that 
it is a perfect fluid with energy-momentum 
tensor
\begin{equation}
T^{\mu \nu} =(e+p)u^{\mu}u^{\nu} - p g^{\mu \nu},
\end{equation}
where $e$ is the energy density and $p$ is the pressure.
This fluid will obey the Euler equation which for 
circular motion can be written as
\begin{equation}
\frac{\partial_{\mu}p}{\rho h}= 
-\partial_{\mu} \ln (u_t) +
\frac{\Omega \partial_{\mu}\ell}{1-\Omega \ell}
=a_{\mu},
\label{KreisBew}
\end{equation}
with $\ell$ being the specific angular momentum defined
in (\ref{eq:ell}), $\rho$ being the mass density and 
$h:=(e+p)/\rho=1+\epsilon +p/\rho$ 
being the specific enthalpy. $\epsilon$ is the density of 
the internal energy. We will consider the case that the fluid 
satisfies a barotropic equation of state. We have already mentioned
above the relativistic von Zeipel theorem which says that
then the differential $d \Omega$ is proportional to the
differential $d \ell$ provided that the latter is non-zero.
Below, however, we will consider the case that $\ell$ is a 
constant, i.e., that $d \ell$ is everywhere equal to zero. 
In this case, the assumption of the von Zeipel theorem does not hold;
however, the  surfaces $\Omega = \mathrm{constant}$ coincide with 
the von Zeipel cylinders $\mathcal{R} = \mathrm{constant}$.
 
The idea of constructing Polish doughnut solutions 
is based on the observation that (\ref{KreisBew}) can
be integrated, assuming a barotropic equation of state 
and making use of the fact that the pressure vanishes 
on the surface of the fluid body. This results in
\begin{equation}
W-W_{\mathrm{in}}:=
-\int_{0}^p\frac{dp'}{\rho h}
=\ln(u_t)-\ln \big((u_t)_{\mathrm{in}} \big)-
\int_{\ell_{\mathrm{in}}}^{\ell}\frac{\Omega d\ell'}{1-\Omega \ell'},
\label{EP-Diff}
\end{equation}
where $W_{\mathrm{in}}$ is a constant of integration. In general,
$\ell$ is a function of $r$ and $\theta$. However, the
simplest idea is to look for solutions where 
$\ell =\mathrm{const}$. In this particular 
case the integral on the right-hand side of (\ref{EP-Diff}) 
vanishes and imposing the condition that $W \to 0$ for 
$r \to \infty$ we obtain in the NUT space-time
\begin{equation}
W=\ln (u_t )
=\frac{1}{2}\ln \left( \frac{\Delta \Sigma \sin^2\theta}{
\Sigma^2\sin^2\theta - \Delta (\ell -2nC-2n\cos\theta)^2 } \right) \, .
\label{EP}
\end{equation}
Here we have used the normalisation condition of the 
four-velocity to express $u_t$ as 
\begin{eqnarray}\label{eq:ut}
u_t=
\sqrt{\frac{g_{t \phi}^2-g_{tt} g_{\phi \phi}}{- g_{\phi \phi} - 2g_{t\phi}\ell - g_{tt}\ell^2}} \, .
\end{eqnarray}
For each value of $\ell$, we have to restrict to the space-time region where the condition
(\ref{eq:elltime-like}) for time-like motion is satisfied. On this region, the von Zeipel cylinders
$\mathcal{R} = \mathrm{constant}$ give the surfaces of constant $\Omega$.
 
Specifying further to a polytropic equation of state
\begin{eqnarray}\label{eq:oly}
p= K \rho^{\Gamma}
\end{eqnarray}
and, correspondingly,
\begin{eqnarray}\label{eq:esilon}
\epsilon=\frac{K \rho^{\Gamma-1}}{\Gamma-1} 
\end{eqnarray}
with $K$ and $\Gamma$ being constants, we can compute the 
pressure integral in (\ref{EP-Diff}) and obtain for the 
density distribution of the matter:
\begin{equation}
\rho(r, \theta )= 
\Bigg( 
\frac{\Gamma -1}{K \Gamma} \bigg( \exp  \Big( W_{\mathrm{in}} -W(r,\theta; \ell ) \Big) -1 \bigg) 
\Bigg) ^{1/(\Gamma -1)} \, .
\label{rho}
\end{equation}
The perfect fluid fills the region where $\rho \ge 0$, i.e., where
$W(r,\theta; \ell )  \le W_{\mathrm{in}}$. 
The construction of Polish doughnut solutions with $\ell =\mathrm{constant}$
depends on two free parameters, the specific angular 
momentum $\ell$ and the integration constant $W_{\mathrm{in}}$. Different 
values one may choose for the constant $\ell$ define different geometrical 
configurations of the rotating matter (see e. g. \cite{Font02}). The constant
$W_{\mathrm{in}}$ determines which of the equipotential surfaces is the
boundary of the fluid. Note that the shape of the Polish doughnut is 
independent of $\Gamma$ and $K$.

The only place where the Manko-Ruiz parameter $C$ enters into this construction
is in the potential (\ref{EP}). We see that changing from $C$ to $C'$ can be
compensated for by changing the constant $\ell$ to $\ell ' = \ell + 2n(C' - C)$. This
demonstrates that it is no restriction of generality if we assume $C=0$ for the
rest of this section.

\begin{figure}
\begin{flushright}
\includegraphics[width=.7\linewidth]{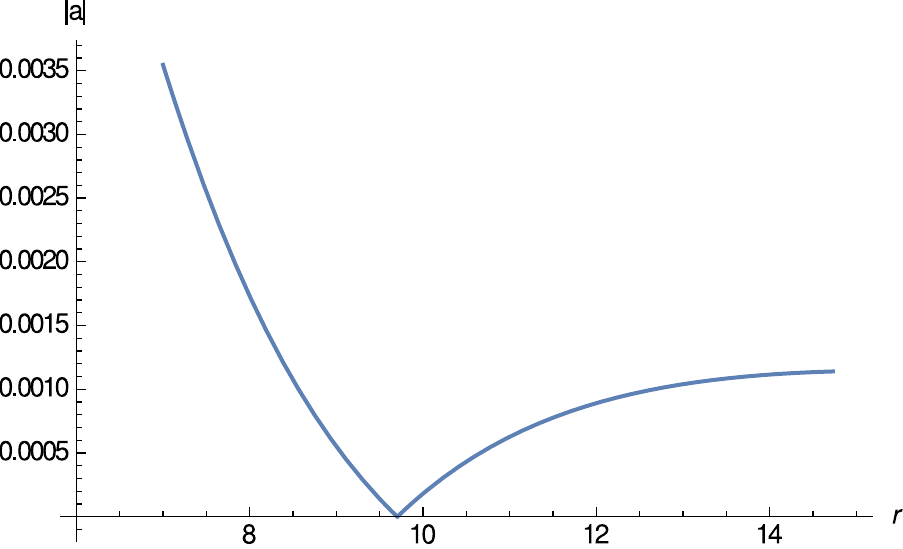}
\end{flushright}
\caption{Modulus $|a|=\sqrt{a _{\mu} a ^{\mu} \begin{matrix} \, \\[-0.1cm] \, \end{matrix} }$ 
of the acceleration $a_{\mu}$ from (\ref{KreisBew}) as a function of $r$ on the surface
of circular geodesic orbits inside a torus with parameters $r_{\mathrm{cen}}=9.7M$, 
$r_{\mathrm{in}}=7M$. The radius coordinate $r$ is given in units of $M$ and the acceleration 
$|a|$ is given in units 
of $M^{-1}$.}
\label{AccDistribution}
\end{figure}

From (\ref{KreisBew}) and (\ref{EP}) we read that our assumption $\partial _{\mu} \ell =0$
implies that $a_{\mu} = - \partial _{\mu} W$, so the motion is geodesic at the critical
points of $W$. We could determine these points by differentiating (\ref{EP}). As an
alternative, we may use the equation $\ell = L/E$ and insert the expressions (\ref{EnCircNUT}) 
and (\ref{DrehimpulsKreisNUT}) for geodesic motion. As we assume $C=0$, this results in 
\begin{equation}\label{eq:ellgeo}
\ell = \frac{2n}{\cos \theta} \, .
\end{equation}
With $\mathrm{cos} ^2 \theta$ given by (\ref{CosNUT}), we get the ``Keplerian'' (i.e.,
geodesic) specific angular momentum as a function of $r$,
\begin{eqnarray}\label{eq:ellK}
\ell _K (r) ^2 = Q(r) \, , \quad r_{\mathrm{ph}} < r < \infty \, ,
\end{eqnarray}
where $Q(r)$ is defined in (\ref{eq:Q}).  Depending on the value of $\ell$ the equation
$\ell = \ell _K (r)$ can have none, one  or two  solutions $r$ outside of the horizon.
Astrophysically interesting configurations occur when there are two such solutions, 
a local minimum and a saddle of $W$. This is the case if 
$\ell _K (r_{\mathrm{ms}})<\ell<\ell _K (r_{\mathrm{mb}})$.
 (Recall that $r_{\mathrm{ms}}$ and $r_{\mathrm{mb}}$ are
determined by (\ref{ISCO-eqLast})  and (\ref{eq:mb}), respectively.)
The first inequality $\ell _K (r_{\mathrm{ms}}) < \ell$ guarantees that $W$ has closed 
equipotential surfaces and a local minimum at a centre with radius coordinate $r_{\mathrm{cen}}$. 
There the density and, thus, the pressure reaches a maximum and the matter moves on a geodesic,
see Fig.~\ref{AccDistribution} for a plot of the acceleration near $r_{\mathrm{cen}}$.
If $\ell _K (r_{\mathrm{ms}})<\ell<\ell _K (r_{\mathrm{mb}})$, the region with closed equipotential surfaces
is bounded by an equipotential surface that forms a ``cusp" at a radius coordinate $r_{\mathrm{c}}$ 
which lies between $r_{\mathrm{hor}}$ and $r_{\mathrm{cen}}$.  The $\theta$ coordinates
of the centre and of the cusp are to be determined by inserting $r_{\mathrm{cen}}$ and 
$r_ {\mathrm{c}}$, respectively, into (\ref{CosNUT}). 

The choice of a value for $W_{\mathrm{in}}$ defines the size of the configuration
and whether or not it has an inner edge, see \cite{Font02} and \cite{Rezzolla13} for more 
details. In the case $\ell _K (r_{\mathrm{ms}})<\ell<\ell _K (r_{\mathrm{mb}})$, where we 
have a centre and a cusp, an inner edge exists if and only if the cusp is outside the fluid 
configuration, i.e., if and only if $W(r_{\mathrm{cen}},\theta _{\mathrm{cen}})
<W_{\mathrm{in}}<W(r_{\mathrm{c}}, \theta _{\mathrm{c}})$. Then the
boundary of the fluid is topologically a torus, as the name Polish ``doughnut'' suggests.
The inner and the outer edge of the torus are determined by the equations
\begin{eqnarray}\label{eq:edges}
W=W_{\mathrm{in}} \, , \quad  \partial _{\theta} W=0 \, .
\end{eqnarray}
Differentiating (\ref{EP}) with respect to $\theta$, assuming $C=0$, demonstrates that 
the second condition is equivalent to (\ref{eq:ellgeo}), i.e., the edges are on the surface 
of time-like circular geodesic orbits (although they are non-geodesic). In particular, the 
coordinates $r_{\mathrm{in}}$ and $\theta _{\mathrm{in}}$ of the inner edge are related 
by (\ref{CosNUT}). It is then equivalent for the toroidal configuration to use 
instead of the parameters $\ell$ and $W_{\mathrm{in}}$ the two radii $r_{\mathrm{cen}}$ 
and $r_{\mathrm{in}}$. If these two radii are given, the values of $\ell$ and $W _{\mathrm{in}}$ 
to be inserted into (\ref{rho}) are $\ell = \ell _K (r_{\mathrm{cen}})$ and 
$W_{\mathrm{in}}=W \big( r_{\mathrm{in}},\theta _{\mathrm{in}} ; \ell_K (r _{\mathrm{cen}} ) \big)$.

The most important new feature of a Polish doughnut in the NUT space-time, in comparison to the 
Schwarzschild or Kerr case, is in the fact that the symmetry with respect to the equatorial plane is
broken. In particular the centre, the cusp and the edges are not in the equatorial plane but rather
in the surface of circular time-like geodesic orbits.

As we have solved the Euler equation on a fixed space-time background, the model
is consistent only if the back-reaction of the fluid onto the space-time geometry is 
negligible. We check that this is true by calculating the total mass of the torus and 
comparing it with the mass of the black hole, and also by calculating the density
and comparing it with a typical curvature invariant.  

The mass $m$ of the torus is computed using the standard generally relativistic expression
\begin{equation}
m= \int_V T_0^0\sqrt{-g} \mathrm{d} V= 2\pi\int\rho(r,\theta)\sqrt{-g}  \mathrm{d}r\mathrm{d}\theta.
\label{mass}
\end{equation}
We have to choose the parameters such that $m$ is small in comparison to $M$.
 
As  a typical curvature invariant we use the Kretschmann scalar which is given in the NUT
space-time by
\begin{eqnarray}
\fl \quad
K_S = R_{\alpha \beta \gamma \delta}R^{\alpha \beta \gamma \delta}
\\
\nonumber
\fl \qquad
= 
\frac{48 \Big( (M^2-n^2)(r^6-15n^2r^4+15n^4r^2-n^6)+
4Mn^2r (3r^4-10n^2r^2+3n^4) \Big)}{(r^2+n^2)^6}
\end{eqnarray}
where $R^{\alpha}{}_{\beta \gamma \delta}$ is the Riemannian curvature tensor. It
can be checked that the Kretschmann scalar is strictly positive outside the horizon.
As a test that the influence of the fluid on the background geometry is negligible, we compare
 at the centre of the torus the square of the density with the Kretschmann scalar.
Demanding that $\rho ^2|_{r_{\mathrm{cen}}} \ll K_S|_{r_{\mathrm{cen}}}$ 
poses restrictions on the remaining free parameters $K$ and $\Gamma$ of the equation of state.
Choosing the gas to be non-relativistic we fix the value of $\Gamma=5/3$. Then the desired
inequality restricts the possible values of $K$ by 
\begin{equation}
K\gg \frac{\Gamma-1}{\Gamma K_S^{(\Gamma-1)/2}}
\bigg( \exp \Big( W_{\mathrm{in}}-W ( r_{\mathrm{cen}},\theta _{\mathrm{cen}} ; \ell ) \Big) -1 \bigg) \, .
\label{K-inequ}
\end{equation}

Assuming the equation of state for a non-relativistic ideal gas, 
$p= \frac{\rho}{\mu} T$, with $\mu$ being the 
mass of a molecule in the gas, one can also find the temperature distribution inside the torus,
\begin{equation}
T=K \rho^{\Gamma-1}\mu \, .
\end{equation}
We have calculated the temperature at the centre of the torus for our numerical 
examples, see Table~\ref{Table m-T}, where we assumed a hydrogen gas, i.e., 
we inserted for $\mu$ the proton mass. We see that the temperature is fairly 
high, but the so-called coolness parameter $\mu c^2 /(kT)$ is still large enough so
that the assumption of a non-relativistic ideal gas may be viewed as acceptable.
Also note that, for our torus which was constructed with $r_{\mathrm{cen}}$ and
$r_{\mathrm{in}}$ given in units of the black-hole mass $M$, the temperature at
the centre is independent of $M$.
 
\begin{figure}
\begin{flushright}
		\includegraphics[scale=0.7]{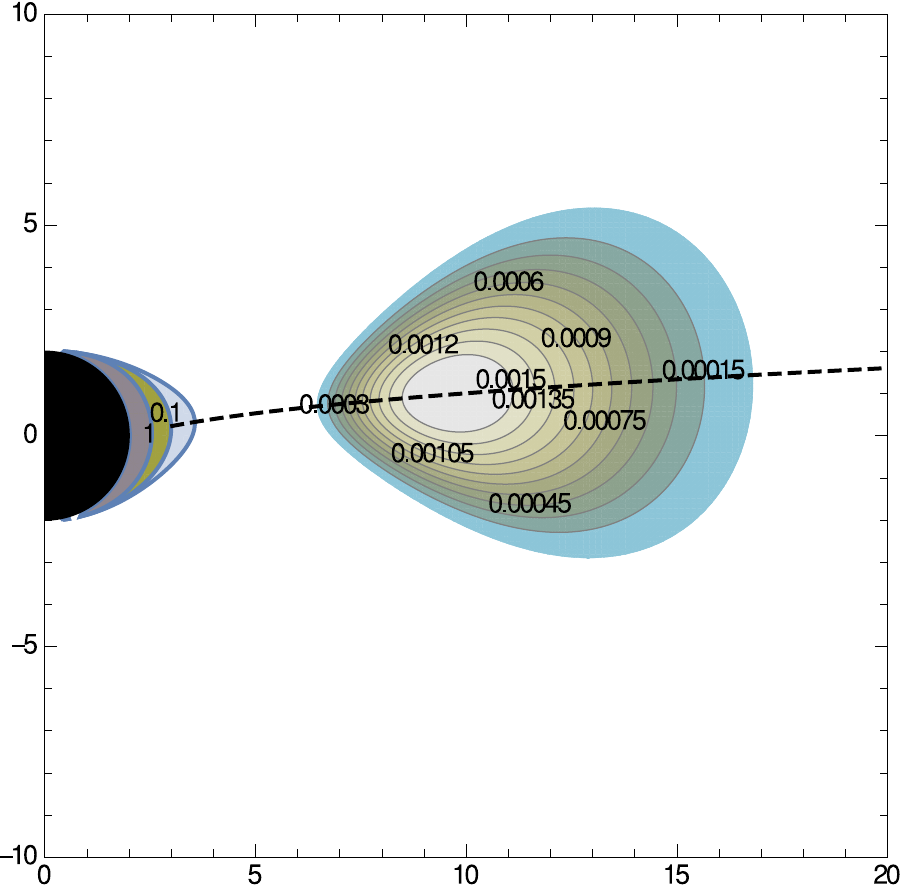}
	\hspace{0.02\linewidth}
		\includegraphics[scale=0.7]{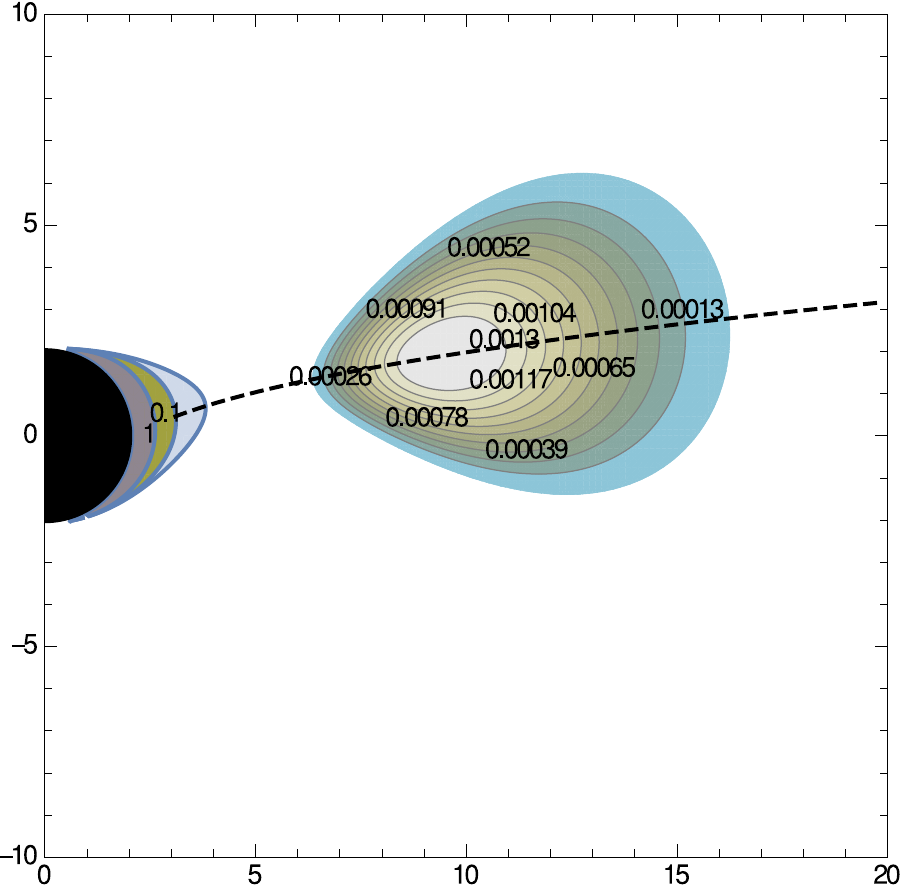}

		\includegraphics[scale=0.7]{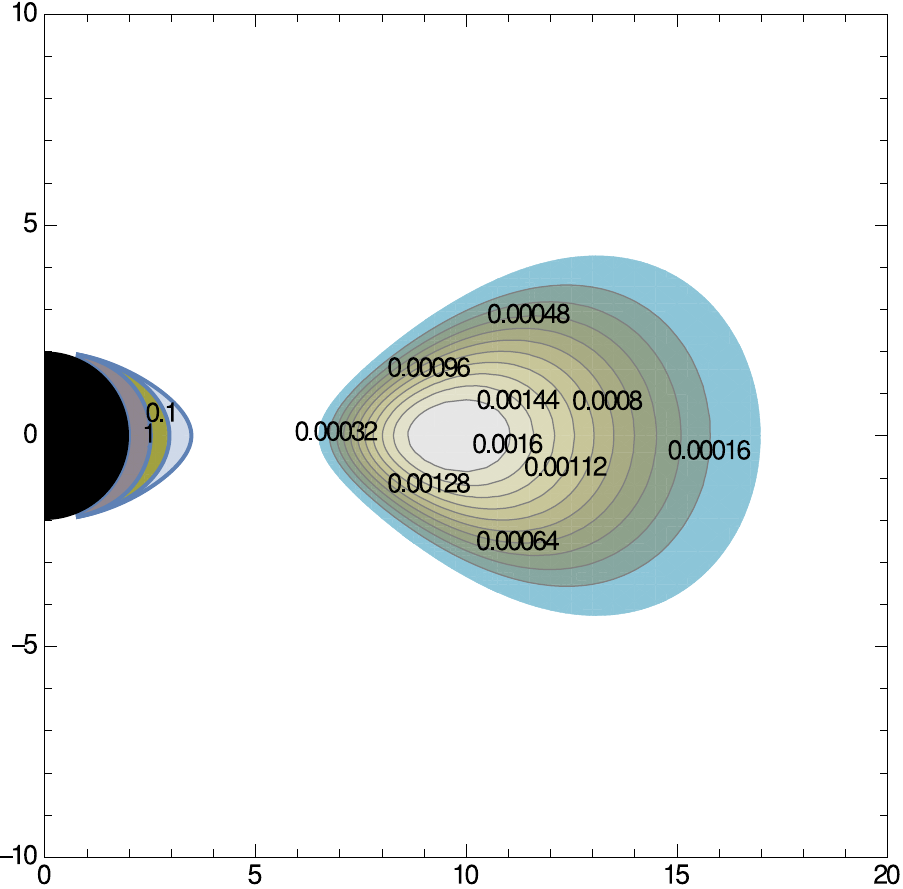}
           \hspace{0.02\linewidth}
		\includegraphics[scale=0.7]{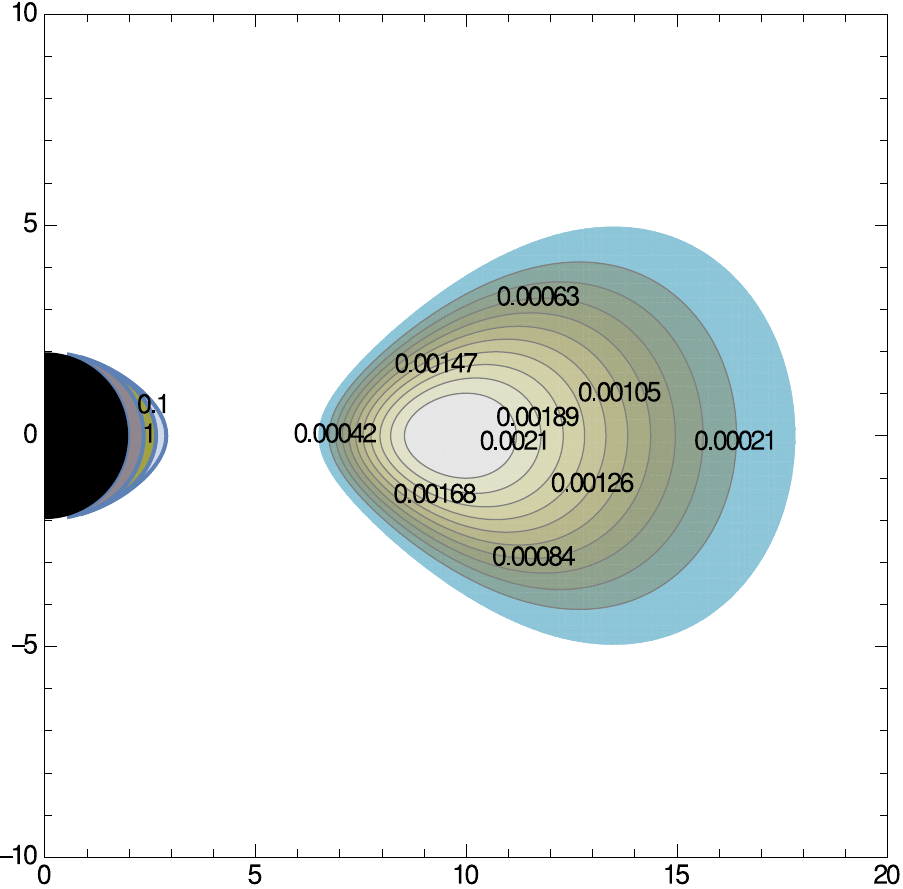}
\end{flushright}
	\caption{Polish Doughnut with $r_{\mathrm{cen}}=9.7 \, M$ and 
	$r_{\mathrm{in}}=6.5 \, M$ in NUT space-time with $n = 0.2 \, M$ (top left),
           in NUT space-time with $n = 0.4 \, M$ (top right), in Schwarzschild space-time
           (bottom left) and in Kerr space-time with $a = 0.2 \, M$ (bottom right).
           As always, we use $M$ as the unit on the axes. Different colours correspond 
	to different values of the matter density $\rho$ given in units of $1/M^2 $.
           A polytropic  equation of state (\ref{eq:oly}) is assumed with $\Gamma = 5/3$ 
           and $K=0.2$ which satisfies the inequality (\ref{K-inequ}) for all the cases
	considered. In the top figures the dashed line marks the surface of circular 
           geodesic orbits. In addition to the torus there is a second region, near the
           horizon, where the mathematical construction gives a positive density. We 
           have included it in the picture but, as explained in the text, it is to be 
           considered as unphysical.} 
	\label{DoughnutNUT}
\end{figure}

In Fig.~\ref{DoughnutNUT} we show Polish doughnuts in NUT space-time for two different values 
of the NUT parameter and, for the sake of comparison, also in Schwarzschild and Kerr space-times.
In all four cases, we have chosen the same values for $r_{\mathrm{in}}$ and $r_{\mathrm{cen}}$.
The total mass of the torus and the density at the centre are given in Table~\ref{Table m-T}. 
The distinguishing feature of a Polish doughnut in NUT space-time is the asymmetry with respect 
to the equatorial plane. The mathematical construction gives a positive density inside the torus and
also in a second region adjacent to the horizon. For the sake of completeness, we have included 
this second region in Fig.~\ref{DoughnutNUT}. However, we do not believe that is has any 
physical relevance because matter on a circular orbit so close to the horizon is expected to
fall into the black hole under the slightest perturbation.   Also, with the chosen parameters
the mass of the fluid inside this region is so big that it could not be neglected in comparison to
the mass of the black hole. For these reasons, we assume that this second fluid configuration is actually not present and we did not include it into the calculation of the configuration mass (\ref{mass}).

\begin{table}
    \begin{flushright}
    \begin{tabular}{|c|c|c|c|c|c|}
    \hline
    $n/M$ & $a/M$& $m/ (10^{-2} M)$ &  $\rho_{\mathrm{cen}}/ (10^{-4}M^{-2})$
     & $T_{\mathrm{cen}} / (10^9 K)$ \\ \hline
    0 & 0 & 5.71 &  2.05 & 7.46 \\ \hline
    0.2 & 0 & 5.28 & 1.99 & 7.01 \\ \hline
    0.4 & 0 & 7.74 &  1.81 & 7.60 \\ \hline
    0 & 0.2 & 7.91 & 2.52 & 8.74 \\ \hline
    \end{tabular}
    \caption{Torus mass $m$, density in the centre $\rho _{\mathrm{cen}}$ and temperature
    in the centre for different values of NUT and Kerr parameters. The values for $n, a, m, 
   \rho_{cen}$ are given in terms of the black-hole mass $M$, the temperature is given in Kelvins.}
    \label{Table m-T}
\end{flushright}
\end{table}

\section*{Conclusions}

In this article we have investigated circular motion of particles in the
domain of outer communication of a NUT black hole. A major difference
in comparison to the case of a Schwarzschild or Kerr black hole is in 
the fact that the geometry is no longer symmetric with respect to the
equatorial plane. This asymmetry is reflected by the fact that 
circular time-like geodesics are not in the equatorial plane but rather 
in a curved surface. This could be observed, e.g. if
a NUT source is surrounded by a thin disc of dust (a ``Saturn ring'').
We have shown that the same asymmetry can be seen in thick
accretion tori known as Polish doughnuts. Our results on the 
surface of circular time-like orbits in three-dimensional  space and on
the shapes of Polish doughnuts apply to Misner's interpretation of
the NUT space-time (with a periodic time coordinate) and equally
well to Bonnor's interpretation (with a singularity on at least one
half-axis). Also, these results are independent of the Manko-Ruiz 
parameter $C$.

For the sake of illustration we have used von Zeipel cylinders with
respect to the stationary observers, $\mathcal{R} = \mathrm{constant}$,
and with respect to the ZAMOs, $\tilde{\mathcal{R}}= \mathrm{constant}$.
Of all the features discussed in this paper, the surfaces $\tilde{\mathcal{R}}
= \mathrm{constant}$ are the only ones that have an essential dependence
on the Manko-Ruiz parameter $C$. The reason is that the ZAMOs depend
on $C$: A (local) coordinate transformation that transforms $C$ to $C'$ 
changes the surfaces $t = \mathrm{constant}$ and thus the ZAMOs.

As we have said in the introduction, the question of whether NUT sources 
exist in Nature is a matter of debate. However, if they do exist, they
could be detected by way of their lensing features, as shown by    
Nouri-Zonoz and Lynden-Bell~\cite{NouriLynden1997}, and also by
way of their influence on orbiting matter, as shown in this paper.

\section*{Acknowledgements}

P. I. Je. was financed through the PhD-student programme Erasmus Mundus Joint Doctorate in International Relativistic Astrophysics (EMJD IRAP) during the course of this work.
Moreover, we gratefully acknowledge support from the DFG 
within the Research Training Group 1620 ``Models of Gravity''. 


\section*{References}

\bibliography{Lib}

\providecommand{\newblock}{}
\begin{thebibliography}{10}
\expandafter\ifx\csname url\endcsname\relax
  \def\url#1{{\tt #1}}\fi
\expandafter\ifx\csname urlprefix\endcsname\relax\def\urlprefix{URL }\fi
\providecommand{\eprint}[2][]{\url{#2}}

\bibitem{NewmanTamburinoUnti1963}
Newman E~T, Tamburino L and Unti T~W~J 1963 {\em J. Math. Phys\/} {\bf 4}
  915--923

\bibitem{BradleyFodorGergelyMarklundPerjes1999}
Bradley M, Fodor G, Gergely L, Marklund M and Perj{\' e}s Z 1999 {\em Class.
  Quantum Grav.\/} {\bf 16} 1667--1675

\bibitem{Taub1951}
Taub A~H 1951 {\em Ann. Math.\/} {\bf 53} 472--490

\bibitem{Misner1963}
Misner C 1963 {\em J. Math. Phys\/} {\bf 4} 924--937

\bibitem{Bonnor1969}
Bonnor W~B 1969 {\em Math. Proc. Cambr. Philos. Soc.\/} {\bf 66} 145--151

\bibitem{MankoRuiz2005}
Manko V~S and Ruiz E 2005 {\em Class. Quantum Grav.\/} {\bf 22} 3555--3560

\bibitem{McGuireRuffini1975}
McGuire P and Ruffini R 1975 {\em Phys. Rev. D\/} {\bf 12} 3026--3029

\bibitem{Misner1967}
Misner C 1967 {\em \emph{in Ehlers J (ed)} Relativity theory and
  astrophysics.I. \emph{(Amer. Math. Soc., Providence, Rhode Island)}\/}  p.
  160

\bibitem{NouriLynden1997}
Nouri-Zonoz M and Lynden-Bell D 1997 {\em Mon. Not. Roy. Soc\/} {\bf 292}
  714--722

\bibitem{Grenzebach14}
Grenzebach A, Perlick V and L{\"a}mmerzahl C 2014 {\em Phys. Rev. D\/} {\bf 89}
  124004

\bibitem{Kagramanova10}
Kagramanova V, Kunz J, Hackmann E and L{\"a}mmerzahl C 2010 {\em Phys. Rev.
  D\/} {\bf 81} 124044

\bibitem{HackmannLaemmerzahl2012}
Hackmann E and L{\"a}mmerzahl C 2012 {\em Phys. Rev. D\/} {\bf 85} 044049

\bibitem{GriffithsPodolsky}
Griffiths J~B and Podolsk{\'y} J 2009 {\em Exact Space-Times in Einstein's
  General Relativity\/} (Cambridge UP)

\bibitem{Chakraborty14}
Chakraborty C 2014 {\em Eur. Phys. J. C\/} {\bf 74} 2759

\bibitem{Abramowicz71}
Abramowicz M~A 1971 {\em Acta Astron.\/} {\bf 21} 81--85

\bibitem{AbramowiczMillerStuchlik1993}
Abramowicz M~A, Miller J~C and Stuchl{\' \i}k Z 1993 {\em Phys. Rev. D\/} {\bf
  47} 1440--1447

\bibitem{Rezzolla13}
Rezzolla L and Zanotti O 2013 {\em Relativistic Hydrodynamics\/} (Oxford
  University Press)

\bibitem{Abramowicz95}
Abramowicz M~A, Nurowski P and Wex N 1995 {\em Class. Quantum Grav.\/} {\bf 12}
  1467--1472

\bibitem{Stuchlik13}
Stuchl\'{i}k Z, Slan\'{y} P, T{\"o}r{\"o}k G and Abramowicz M~A 2005 {\em Phys.
  Rev. D\/} {\bf 71} 024037

\bibitem{Foertsch03}
Foertsch T, Hasse W and Perlick V 2003 {\em Class. Quantum Grav.\/} {\bf 20}
  4635--4651

\bibitem{HassePerlick2006}
Hasse W and Perlick V 2006 {\em J. Math. Phys.\/} {\bf 47} 042503

\bibitem{JaroszynskiAbramowiczPaczynski1980}
Jaroszy{\'n}ski M, Abramowicz M~A and Paczy{\'n}ski B 1980 {\em Acta Astron.\/}
  {\bf 30} 1--34

\bibitem{Font02}
Font J~A and Daigne F 2002 {\em Mon. Not. R. Astron. Soc.\/} {\bf 334} 383--400

\end{thebibliography}
\bibliographystyle{iopart-num}

\end{document}